\documentclass[fleqn,usenatbib]{mnras}

\usepackage{newtxtext,newtxmath}
\usepackage{graphicx}	

\usepackage[T1]{fontenc}

\DeclareRobustCommand{\VAN}[3]{#2}
\let\VANthebibliography\thebibliography
\def\thebibliography{\DeclareRobustCommand{\VAN}[3]{##3}\VANthebibliography}

\title[Modelling stellar convective transport with plumes I]{Modelling stellar convective transport with plumes: I. Non-equilibrium turbulence effect in double-averaging formulation}

\author[N. Yokoi, Y. Masada \& T. Takiwaki]{
N. Yokoi,$^{1}$\thanks{E-mail: nobyokoi@iis.u-tokyo.ac.jp, Visiting researcher at the Nordic Institute for Theoretical Physics (NORDITA)}
Y. Masada,$^{2}$
and T. Takiwaki$^{3}$
\\
$^{1}$Institute of Industrial Science, University of Tokyo, Komaba, Meguro, Tokyo 153-8505, Japan\\
$^{2}$Department of Applied Physics, Faculty of Science, Fukuoka University, Fukuoka 814-0180, Japan\\
$^{3}$Division of Science, National Astronomical Observatory of Japan (NAOJ), Osawa, Mitaka, Tokyo 181-8588, Japan
}

\date{Accepted 20 April 2022. Received 20 March 2022; in original form 17 November 2021}

\pubyear{2022}

\makeatletter
\newenvironment{Eqnarray}%
	{\arraycolsep 0.14em\eqnarray}{\endeqnarray}
\makeatother

\begin{document}
\label{firstpage}
\pagerange{\pageref{firstpage}--\pageref{lastpage}}
\maketitle

\begin{abstract}Plumes in a convective flow are considered to be relevant to the turbulent transport in convection. The effective mass, momentum, and heat transports in the convective turbulence are investigated in the framework of time--space double averaging procedure, where a field quantity is decomposed into three parts: the spatiotemporal mean (spatial average of the time-averaged) field, the dispersion or coherent fluctuation, and the random or incoherent fluctuation. With this framework, turbulent correlations in the mean-field equations are divided into the dispersion/coherent and random/incoherent correlation part. By reckoning the plume as the coherent fluctuation, a transport model for the convective turbulence is constructed with the aid of the non-equilibrium effect, in which the change of turbulence characteristics along the mean stream is taken into account for the modelling of the turbulent transport coefficients. In this work, for the first time, change of turbulence properties along plume motions is incorporated into the expression of the turbulent transport coefficients. This non-equilibrium model is applied to a stellar convective flow. One of the prominent characteristics of a surface cooling-driven convection, the enhanced and localised turbulent mass flux below the surface layer, which cannot be reproduced at all by the usual eddy-diffusivity model with mixing length theory (MLT), is well reproduced by the present model. Our results show that the incorporation of plume motion into turbulent transport model is an important and very relevant extension of mean-field theory beyond the heuristic gradient transport model with MLT.
\end{abstract}

\begin{keywords}
convection -- hydrodynamics -- turbulence -- stars: interiors -- Sun: interior
\end{keywords}



\section{Introduction}\label{sec:intro}
Fluid motions in the stellar convection zone and the planetary atmosphere are convective turbulence driven by a buoyancy force associated with temperature gradient and/or stratified density. Convection in stellar interiors is vigorously turbulent, and plays a crucial role in the energy transport and the magnetic-field generation (dynamos) in the star. Theoretical analysis and modelling of the convective turbulence are indispensable for our deeper understanding of the astrophysical and geophysical flow phenomena. In convective flows, persistent flow structures are ubiquitously observed in local domains in space and time. Jets, plumes, and thermals (plumes are jets driven by buoyancy only, and thermals denote suddenly released buoyant elements mainly in meteorological context) are typical examples of such persistent flow structures. Plumes play a key role in determining the effective transport of the mass, momentum, heat in convective turbulence. For modelling the plume effects, it is known that the entrainment assumptions (or equivalent similarity arguments) can be used \citep{tur1973,lin2000}.

	In the stellar convection studies, it was recognised that a strong downward directed flow plays a key role in dynamics and turbulent transports. Reviewing the experimental and numerical results, it was pointed out that the observed patterns in the stellar convection are dominantly determined by the cooling at the surface \citep{spr1997}. The downward diving plumes, originated at the cooling surface layer, are able to reach the bottom of the convection zone and to contribute to turbulent transport \citep{rie1995}. To construct an elaborated model of stellar convective flow, the effects of diving plumes should be properly taken into account beyond the hydrostatic pressure distribution and the simple velocity-proportional entrainment assumption \citep{ras1998}.

	In the mean-field turbulence model, the evolutions of mean fields are determined by the effective transport represented by the turbulent fluxes such as the turbulent mass flux, the Reynolds stress, the turbulent heat flux, etc. Traditionally, mixing-length theory (MLT) has been employed for describing the convective energy transport in the interior of stars. In the traditional model, the turbulent fluxes are approximated by the gradient-diffusion-type formula with MLT model for the transport coefficients \citep{boe1958,sti2002}. Numerical simulations of stellar convection revealed that the mean-field turbulence models with MLT need to be modified or replaced by a more elaborated formulation including the turbulent cascade by Kolmogorov theory and chaotic behaviour of an integral scale roll of Lorenz \citep{arn2015}. In particular, the non-local transport mechanisms associated with plumes should be implemented into the turbulence model \citep{mur2011,bra2016}. In addition, it was pointed out that the mixing by the downdraft motions driven immediately below the surface of radiative cooling is much more effective than the counterpart by the flux due to the weakly super-adiabatically driven gradient diffusion across the whole convection zone \citep{cos2016}, although recent numerical simulations from base to surface suggest another interpretation \citep{hot2019}. Beyond the simplest gradient-transport-type models, a transport model incorporating the effects of plumes as coherent structures should be constructed \citep{bra2016}. At the same time, as long as the values of the physical parameters in a numerical simulation are far from the realistic ones in the stellar convective turbulence, the interpretation of the simulation results should be done with caution. For example, a recent numerical simulation has revealed a strong dependence of convective overshooting and energy flux on the molecular Prandtl number \citep{kap2021}. In this sense, convection in the Sun is quite different from that obtained from simulations in which $Pr \sim 1$.

	Because of the vast range of scales that must be included, direct numerical simulation of stellar convective flow is simply impossible in the foreseeable future, even using sophisticated algorithms optimised for massively parallel computers. For this reason, developing sophisticated theories and modelling of realistic turbulence is indispensable for the study of stellar convection. Several critical deficiencies of the simple eddy-viscosity representation have been clarified. One deficiency is the lack of vorticity effect. An alleviation of this deficiency was proposed by the inclusion of helicity effect coupled with the mean vorticity and/or rotation \citep{yok1993,yok2016}. Another possible way to alleviate the drawback of a turbulence model using the usual gradient-diffusion approximation is to modify the model by incorporating the non-equilibrium effect into the expressions for the turbulent fluxes. Variations of the turbulence characteristics in time or along the mean stream can be taken into account as a non-equilibrium effect on turbulent transport \citep{yos1993,yok2022}. The presentation of the non-equilibrium effect itself may take on various aspects. Beyond the entrainment assumptions, the non-equilibrium or time-dependent effect has been needed in modelling cloud dynamics in a non-uniform environment [for example, see Chap.~6 in \citet{tur1973}]. Here in this work, we focus our arguments on the non-equilibrium effect associated with variations of turbulence along the advective motion \citep{yos1994}. In the presence of non-equilibrium variation of the turbulent energy and its dissipation rate, the time and length scales of the turbulence are altered. Such non-equilibrium properties of turbulence should affect the model expression of the turbulent transport. As the multiple-scale direct-interaction approximation, an analytical theory for strongly non-linear and inhomogeneous turbulence, shows, the gradient-diffusion-type model for the turbulent fluxes is closely linked to the equilibrium property of turbulence statistics. Inclusion of the non-equilibrium properties of turbulence statistics leads to a deviation from the gradient-diffusion-type model for the turbulent transports (see later in \S~\ref{sec:non-equiv}). Since the non-equilibrium effect stems from the variation of the turbulent statistics along the advective motion, some kind of convective flows such as jets, thermals, and plumes can be argued in the context of the non-equilibrium effect. This non-equilibrium property of the coherent jets and plumes in laboratory experiments has been recently discussed \citep{sun2021,yok2022}.

	However, we face some difficulty in applying the non-equilibrium effect formulation to convective turbulent flows. In the simplest formulation of the non-equilibrium effect, the effect is represented by the material derivative based on the mean velocity, $D/Dt \equiv \partial / \partial t + {\bf{U}} \cdot \nabla$ (${\bf{U}}$: mean velocity). In the case of the closed-domain convective motions such as the flow in stellar convection zone and the Rayleigh--B\'{e}nard convection, which have been studied in detail in experimental and numerical manners, the mean velocity under the simple ensemble averaging or space averaging over the horizontal surface in the homogeneous directions is typically negligibly small (${\bf{U}} \simeq 0$) because of the statistical smearing out and it is not suitable for representing the local velocity structure such as plumes. These spatiotemporal structures (plumes, convective flows, jets, etc.) certainly exist locally in time and space, but will disappear under a simple averaging procedure such as the ensemble, space and time averaging. In the sense that the average is zero, these flow structures belong to the fluctuation, but should be treated as the coherent or structural component of the fluctuation. For the purpose of incorporating the plume effects into a turbulence model for convective flows, we adopt a time--space double averaging method, a formulation that can contrast the coherent/structural fluctuation component with the incoherent/random fluctuation one. 
	
	This formulation is to be applied to a flow configuration relevant to stellar convection. If the convective motion is cooling-driven at the near surface layer, the turbulent transport is dominated by the cool diving plume. As will be shown in \S~\ref{sec:stell_conv}, the property of turbulence transport in the non-local convection vigorously driven by cooling at the surface is fairly different from the one in the local convection driven by weakly superadiabatic ambient state across the full depth. For instance, the turbulent mass flux $\langle {\rho' {\bf{u}}'} \rangle$ is much larger in the cooling-driven case, and the peak of the flux is located in the shallow region ($\rho'$: density fluctuation, ${\bf{u}}'$: velocity fluctuation, $\langle \cdots \rangle$: mean or averaging). As mentioned above, unlike the turbulent transport in the weakly superadiabatic throughout the convection zone case (the local transport case), the turbulent transport in the cooling-driven convection case (non-local transport case) cannot be properly described by the gradient-transport type model, and the turbulence model based on a simple mixing-length theory (MLT) should be modified for the cooling-driven convection \citep{cos2016,bra2016}. We implement the non-equilibrium effect into the convection turbulence model in the framework of the time--space double averaging, and apply this model to the cooling-driven convection.

	Along this line of thought, we are preparing two papers on this subject. The first one (Paper I) is the present paper, which mainly focuses on the theoretical and analytical framework of the turbulence modelling with the non-equilibrium effect. For the purpose of capturing the plume motions, the basic notions of space--time double averaging procedure as well as the evolution equations of the coherent and incoherent fluctuation stresses and energies are presented in Paper I. In addition, with the aid of the direct numerical simulations (DNSs), the basic validation of the non-equilibrium turbulence model is presented in the context of the stellar convection. In the second paper (Paper II), we will present the details of the model setup, numerical results, and data analysis. The contents of Paper II include the detailed numerical results on the Fourier spectra and probability distribution function (PDF) of convection velocity, turbulent mass, momentum and energy transports, as well as the data analysis methods with the Fourier filtering and double averaging \citep{mas2022}.

	The organisation of this paper (Paper I) is as follows. The fundamental equations as well as the mean-field equations with several turbulent fluxes are presented in \S~\ref{sec:m-field_turb_correl}. After presenting the basic notions of the double-averaging procedure and its property in \S~\ref{sec:ave_methods}, the evolution equations of some turbulence correlations and energies are given in \S~\ref{sec:eqs_turb_correl}, with special emphasis on the interaction between the coherent and incoherent fluctuation motions. In order to incorporate the non-equilibrium effect into the stellar convection model, in \S~\ref{sec:model_conv}, the plume effects are viewed from the double-averaging procedure. The model structure in the double-averaging methodology is also examined. In \S\ref{sec:stell_conv}, the model is applied to a flow configuration relevant to the stellar convection to describe the spatial distribution of the turbulent mass flux in the local and non-local convection cases. Conclusions are presented in \S~\ref{sec:concl}.

\section{Fundamental equations and turbulent correlations}
\label{sec:m-field_turb_correl}
The system of equations for the compressible hydrodynamic flow with the external force included can be written as

\begin{equation}
	\frac{\partial \rho}{\partial t}
	+ \nabla \cdot (\rho {\bf{u}})
	= 0,
	\label{eq:den_eq}
\end{equation}
\begin{equation}
	\frac{\partial}{\partial t} \rho u^i
	+ \frac{\partial}{\partial x^j} \rho u^j u^i
	= - \frac{\partial p}{\partial x^i}
	+ \frac{\partial}{\partial x^j} \mu s^{ji}
	+ f_{\rm{ex}}^i,
	\label{eq:vel_eq}
\end{equation}
\begin{equation}
	\frac{\partial}{\partial t} \rho e
	+ \nabla \cdot \left( {\rho {\bf{u}} e} \right)
	= \nabla \cdot \left( {\eta \nabla\theta} \right)
	- p \nabla \cdot {\bf{u}}
	+ \phi
	+ \zeta,
	\label{eq:int_en_eq}
\end{equation}
where $\rho$ is the density, ${\bf{u}}$ the velocity, $p$ the pressure, $e$ the internal energy, $\mu$ the viscosity, $\eta$ the thermal diffusivity, $s^{ij}$ the deviatoric or traceless part of the velocity strain defined by
\begin{equation}
	s^{ij}
	= \frac{\partial u^j}{\partial x^i}
	+ \frac{\partial u^i}{\partial x^j}
	- \frac{2}{3} \nabla \cdot {\bf{u}} \delta^{ij}.
	\label{eq:vel_strain_def}
\end{equation}
In (\ref{eq:vel_eq}), ${\bf{f}}_{\rm{ex}}$ is the external force. In the buoyantly convective flow, we consider the force of gravity for ${\bf{f}}_{\rm{ex}}$: 
\begin{equation}
	{\bf{f}}_{\rm{ex}} = \rho {\bf{g}},
	\label{eq:gravity_force}
\end{equation}
where ${\bf{g}}$ is the acceleration due to gravity. In (\ref{eq:int_en_eq}), $\phi$ is the dissipation function that represents the conversion of the kinetic energy to heat through the viscosity effect:
\begin{equation}
	\phi = \mu s^{ij} \frac{\partial u^i}{\partial x^j},
	\label{eq:diss_fn}
\end{equation}
and $\zeta$ is the internal energy source/sink term.

	The pressure $p$ is related to the temperature $\theta$ and the internal energy $e$ as
\begin{equation}
	p = R \rho \theta = (\gamma - 1) \rho e,
	\label{eq:eos}
\end{equation}
where
\begin{equation}
	e = C_V(\theta) \theta.
	\label{eq:e-theta_rel}
\end{equation}
Here, $C_V$ is the specific heat at constant volume, $R$ is the gas constant, and $\gamma$ is the ratio of $C_P$ (the specific heat at constant pressure) to $C_V$.

	We first adopt the simple decomposition of a field quantity $f$ into the mean $\langle {f} \rangle (\equiv F)$ and the fluctuation around it, $f'$, as
\begin{equation}
	f = F + f',\;\;\; F = \langle {f} \rangle
	\label{eq:mean_fluct_decomp}
\end{equation}
with
\begin{subequations}\label{eq:f_F_f'}
\begin{equation}
	f = (\rho, {\bf{u}}, p, e, \theta),
	\label{eq:fld_qnts}
\end{equation}
\begin{equation}
	F = (\langle {\rho} \rangle, {\bf{U}}, P, E, \Theta),
	\label{eq:mean_fld_qnts}
\end{equation}
\begin{equation}
	f' = (\rho', {\bf{u}}', p', e', \theta')
	\label{eq:fluct_fld_qnts}
\end{equation}
\end{subequations}
($\langle \cdot \rangle$: ensemble average or space average in the homogeneous directions). Under this decomposition, the mean-field equations are given as
\begin{equation}
	\frac{\partial \langle {\rho} \rangle}{\partial t}
	+ \nabla \cdot \left( {
		\langle {\rho} \rangle {\bf{U}}
	} \right)
	= - \nabla \cdot \langle {\rho' {\bf{u}}'} \rangle,
	\label{eq:mean_den_eq}
\end{equation}
\begin{eqnarray}
	\lefteqn{
	\frac{\partial}{\partial t} 
		\langle {\rho} \rangle U^i
	+ \frac{\partial}{\partial x^j} \langle{\rho}\rangle U^j U^i
	}\nonumber\\
	&&\hspace{-10pt}= - (\gamma - 1) 
    \frac{\partial}{\partial x^i} 
		\langle {\rho} \rangle E
	+ \frac{\partial}{\partial x^j} {\mu} {\cal{S}}^{ji}
	\nonumber\\
	&&\hspace{-10pt}- \frac{\partial}{\partial x^j} \left( {
		\langle {\rho} \rangle 
		\langle {u'{}^j u'{}^i} \rangle
		\rule{0.ex}{3.ex}
		+ U^j \langle {\rho' u'{}^i} \rangle
		+ U^i \langle {\rho' u'{}^j} \rangle
	} \right),
	\label{eq:mean_mom_eq}
\end{eqnarray}
\begin{eqnarray}
	\lefteqn{
	\frac{\partial}{\partial t} 
		\langle {\rho} \rangle E
	+ \nabla \cdot (\langle {\rho} \rangle {\bf{U}} E)
	}\nonumber\\
	&&\hspace{-17pt} = \nabla \cdot \left( {
		\frac{\kappa}{C_v} \nabla E
	} \right)
	- \nabla \cdot \left( { \rule{0.ex}{3.ex}
		\langle {\rho} \rangle \langle {e' {\bf{u}}'} \rangle
		+ E \langle {\rho' {\bf{u}}'} \rangle
		+ {\bf{U}} \langle {\rho' e'} \rangle
	} \right)
	\nonumber\\
	&&\hspace{-17pt} - (\gamma - 1) \left( { \rule{0.ex}{3.ex}
		\langle {\rho} \rangle E \nabla \cdot {\bf{U}}
	+ \langle {\rho} \rangle 
		\langle {e' \nabla \cdot {\textbf{u}}'} \rangle
	+ E \langle {\rho' \nabla \cdot {\bf{u}}'} \rangle
	} \right),
	\label{eq:mean_int_en_eq}
\end{eqnarray}
where ${\cal{S}}^{ij}$ in (\ref{eq:mean_mom_eq}) is the mean-velocity counterpart of the strain rate (\ref{eq:vel_strain_def}), defined by
\begin{equation}
	{\cal{S}}^{ij}
	= \frac{\partial U^j}{\partial x^i}
	+ \frac{\partial U^i}{\partial x^j}
	- \frac{2}{3} \nabla \cdot {\bf{U}}\delta^{ij}.
	\label{eq:mean_strain_rate}
\end{equation}

	The turbulent correlations in (\ref{eq:mean_den_eq})-(\ref{eq:mean_int_en_eq}), the turbulent mass flux $\langle {\rho' {\bf{u}}'} \rangle$, the Reynolds stress $\langle {u'{}^i u'{}^j} \rangle$, the turbulent internal-energy flux $\langle {e' {\bf{u}}'} \rangle$, etc.\ are the most important quantities, which determine the transport in the mean-field equations due to turbulence. The expressions of these turbulent fluxes should be obtained from the equations of the fluctuating density $\rho'$, velocity ${\bf{u}}'$ and internal energy $e'$. The fluctuation-field equations are given as
\begin{Eqnarray}
	\frac{\partial \rho'}{\partial t}
	&+& \left( {{\bf{U}} \cdot \nabla} \right) \rho'
		+ \nabla \cdot \left( {\rho' {\bf{u}}'} \right)
		+ \langle {\rho} \rangle \nabla \cdot {\bf{u}}'
	\nonumber\\
	&=& - \left( {{\bf{u}}' \cdot \nabla} \right) 
		\langle {\rho} \rangle
		- \rho' \nabla \cdot {\bf{U}}
		+ R_{\rho},
	\label{eq:fluct_rho_eq}
\end{Eqnarray}
\begin{Eqnarray}
	\frac{Du'{}^i}{Dt}
	&=& - \left( {{\bf{u}}' \cdot \nabla} \right) u'{}^i
	+ \frac{1}{\langle {\rho} \rangle}
	\frac{\partial}{\partial x^j} \mu s'{}^{ji}
	\nonumber\\
	&&\hspace{-20pt} - \left( {\gamma - 1} \right) \left( {
		\frac{\partial e'}{\partial x^i}
		+ \frac{E}{\langle {\rho} \rangle} 
			\frac{\partial \rho'}{\partial x^i}
	} \right)
	- ({\bf{u}}' \cdot \nabla) \langle {u} \rangle^i
	\nonumber\\
	&&\hspace{-20pt}- (\gamma - 1) \left( {
		\frac{\rho'}{\langle {\rho} \rangle} 
		\frac{\partial E}{\partial x^i}
		+ \frac{e'}{\langle {\rho} \rangle} 
			\frac{\partial \langle {\rho} \rangle}{\partial x^i}
		} \right)
	- \frac{\rho'}{\langle {\rho} \rangle} 
		\frac{DU^i}{Dt}
	+ R_{u}^i,
	\label{eq:fluct_vel_eq}
\end{Eqnarray}
\begin{Eqnarray}
	\frac{\partial e'}{\partial t}
	&+& ({\bf{U}} \cdot \nabla) e'
	= - ({\bf{u}}' \cdot \nabla) e'
		+ \frac{1}{\langle {\rho} \rangle} \nabla \cdot 
			\left( {\frac{\kappa}{C_v} \nabla e'} \right)
	\nonumber\\
	&-& (\gamma - 1) E \nabla \cdot {\bf{u}}'
		- ({\bf{u}}' \cdot \nabla) E
	\nonumber\\
	&-& (\gamma - 1) \left( {
		e' + \frac{\rho'}{\langle {\rho} \rangle} E
		} \right) \nabla \cdot {\bf{U}}
	+ R_e,
	\label{eq:fluct_int_en_eq}
\end{Eqnarray}
where $s'{}^{ij}$ is the strain rate of the fluctuating velocity defined by
\begin{equation}
	s'{}^{ij}
	= \frac{\partial u'{}^j}{\partial x^i}
	+ \frac{\partial u'{}^i}{\partial x^j}
	- \frac{2}{3} \nabla \cdot {\bf{u}}' \delta^{ij}.
	\label{eq:fluct_strain_rate}
\end{equation}
Here, $R_\rho$, $R_u$, and $R_e$ represent the residual terms consisting of the turbulent correlations like $\langle {\rho' {\bf{u}}'} \rangle$, $\langle {u'{}^i u'{}^j} \rangle$, $\langle {e' {\bf{u}}'} \rangle$, etc.\ and higher-order correlations. Note that these fluctuation-field equations as well as the mean-field equations are derived from the fundamental equations. So, both the mean- and fluctuation-field equations contain the turbulent fluxes [see textbooks, for instance, \citet{mat2000,yok2020}]. However, the details of the residual terms are suppressed here since they do not contribute to the later discussion on the generation mechanisms of fluctuations. 

	With the aid of the two-scale direct-interaction approximation (TSDIA), a multiple-scale renormalisation perturbation expansion, the turbulent correlations are expressed in terms of the spectral and response functions of turbulence \citep{yos1984,yok2020}. The analytical expressions  of the turbulent correlations, and the corresponding model expressions based on the theory are given in (\ref{eq:model_uu_exp})-(\ref{eq:model_rho_dil_exp}) in Appendix~\ref{sec:append_A} [the hydrodynamic limit of \citet{yok2018a,yok2018b}].   

	In the turbulence modelling approach, turbulent transport coefficients in (\ref{eq:model_uu_exp})-(\ref{eq:model_rho_dil_exp}), such as the eddy viscosity $\nu_{\rm{T}}$, turbulent diffusivity $\kappa_{\rho}$, turbulent internal-energy diffusivity $\eta_{E}$, etc., should reflect the statistical properties of the turbulence in consideration. In a self-consistent turbulence model, where the mean and turbulent fields are simultaneously and consistently determined by the nonlinear dynamics of turbulent flow without resorting to externally determined transport coefficients, the expressions of the transport coefficients have to be expressed in terms of a few statistical quantities that properly represent the nonlinear dynamics of turbulence. A possible way to choose such turbulent statistical quantities in compressible turbulent flows is choosing the turbulent energy (per mass) $K$, its dissipation rate $\varepsilon$, and the density variance $K_\rho$. They are defined by
\begin{equation}
	K = \langle {{\bf{u}}'{}^2} \rangle/2,
	\label{eq:K_def}
\end{equation}
\begin{equation}
	\varepsilon = \nu \left\langle {
		 \frac{\partial u'{}^j}{\partial x^i} s'{}^{ij} 
	} \right\rangle,
	\label{eq:eps_def}
\end{equation}
\begin{equation}
	K_\rho = \langle {\rho'{}^2} \rangle.
	\label {eq:K_rho_def}
\end{equation}
The evolution equations of these turbulent statistical quantities are obtained from the equations of the fluctuating fields (\ref{eq:fluct_rho_eq})-(\ref{eq:fluct_int_en_eq}). The evolution equations of $K$, $\varepsilon$, and $K_\rho$ are given in (\ref{eq:model_K_eq})-(\ref{eq:model_den_var_eq}) in Appendix~\ref{sec:append_A}.

\section{Averaging methods}\label{sec:ave_methods}
\subsection{Conditional averaging}\label{sec:cond_ave}
There are several ways to extract the local spatiotemporal structures from the random fluctuations. The conditional averaging procedure is one of such ways. For example in the turbulent boundary-layer study, we consider the streamwise and wall-normal velocity fluctuation components, $u'$ and $v'$, and divide the whole value domain of fluctuations into the four quadrants depending on the signs of $(u', v') = (+,+), (-,+), (-,-), (+,-)$, and examine the statistical properties in each quadrant. For example, the motions belonging to the quadrant $(+,+)$ (or $(-,-)$) represents an event at which the turbulent flow is moving along (or opposite to) the streamwise direction while departing (approaching) from the wall. In this sense, the statistics based on the four quadrants should correspond to a conditional averaging linked to the type of events, such as the bursting, sweeping, etc. in the turbulent boundary layer [see \citet{wal2013} and references cited therein]. For the convective turbulence with plumes, a conditional averaging within the four quadrants based on the combination of the vertical component of the velocity fluctuation, $w'$, and the temperature fluctuation $\theta'$, $(w', \theta')$ is possible. In this case, for example, the motions with $(w',\theta') = (+,+)$ represents an ascending plume with being heated, and $(-,-)$ does a descending plume with being cooled.

\subsection{Double averaging method}\label{sec:double_ave}
Another way to represent the coherent and incoherent fluctuations is to adopt a double-averaging methodology \citep{fin2008,pok2008,dey2014,dey2020}. We regard convection plumes as a turbulent coherent fluctuation, and investigate the properties of the fluctuations. There are several types of the double averaging method. The representative double-averaging procedure is the double filtering ubiquitously adopted in the data processing and the large-eddy simulations (LESs) of turbulent flows. By a sequential application of two or more filters with varied filtering levels (in frequency or length scale) to the raw data, slowly varying fluctuating motions are extracted from the fast varying fluctuating ones \citep{sag2006}. 

	In this work, we consider a combination of the time and space averaging. Depending on the sequential order of time and space averaging, there are two ways of the time and space double averaging method: (i) the time--space averaging, where the spatial averaging is taken to the already temporary-averaged variables; and (ii) the space--time averaging, where the temporal averaging is taken to the already spatial averaged variables. The first one, the time--space averaging is more appropriate for the purpose of extracting the plume structures. As is usual for the case of double averaging or filtering procedures, the averaging should be taken first with a finer resolution manner, then a ``coarse grained'' averaging should be taken. Otherwise, the coarse grained averaging makes the fine structures be smeared out. In this work, the averaging window for the time average is set much shorter than the counterpart for the space average.\footnote{The fraction of the lifetime of a plume, $\tilde{\tau}$, to the averaging time window $T$ is defined by $\Delta_{\rm{time}} = \tilde{\tau}/ T$, while the fraction of the occupation domain area of a plume, $\tilde{s} = b^2$ ($b$: horizontal dimension of the plume), to the space averaging window $\Delta S = \Delta_x \Delta_y$ is $\Delta_{\rm{space}} = \tilde{s} / \Delta S$. If these fractions are much smaller than unity, the plume effects are statistically smeared out by the averaging. On the other hand, if the fraction is comparable to unity, the plume effects are detectable after the averaging. Because of the spatial localisation of a plume, $\Delta_{\rm{space}}$ is very small ($\Delta_{\rm{space}} \ll 1$), whereas $\Delta_{\rm{time}}$ can be comparable to unity if we set the averaging time window $T$ similar to the lifetime $\tilde{\tau}$ ($T \simeq \tilde{\tau}$). This is the reason why we should take the time averaging first.} So, we should first take the temporal average over a time period during which a plume persistently exists, then perform the space average of the already time-averaged quantities. We further assume that the space averaging provide a surrogate of the ensemble averaging.

	As for the time average, the usual time average defined by
\begin{equation}
	\overline{f}({\textbf{r}})
	= \lim_{T \to \infty} \frac{1}{T} \int_{0}^{T} f({\textbf{r}};t) dt
	\label{eq:time_ave_gen}
\end{equation}
is not suitable at all for detecting the plume structure since their structures are smeared out during the long averaging time $T \to \infty$. 
In order to catch a plume structure, which is local in time and space, we adopt a time filter defined by
\begin{equation}
	\overline{f}({\textbf{r}};t)
	= \frac{1}{T} \int_{t-T/2}^{t+T/2} f({\textbf{r}};s) ds.
	\label{eq:time_ave_short}
\end{equation}
Here, as for the appropriate averaging time $T$, we adopt a time which is much longer than the eddy turn-over time of turbulence, $\tau$, and much shorter than the time scale of the mean-field evolution, $\Xi$, as
\begin{equation}
	\tau \ll T \ll \Xi.
	\label{eq:ave_time_cond}
\end{equation}
Of course, all the statistics depend on the value of the averaging time $T$. It is difficult to determine the averaging time $T$ in a general manner. Here, we adopt $T$ which is similar to the plume lifetime (time of the persistent presence of a plume). The eddy turn-over time of the turbulence $\tau$ and the time scale of the mean-field evolution $\Xi$ are determined by the density scale height, intensity of turbulence, the depth of the convection zone, etc. As will be discussed at the end of \S~\ref{sec:incoh_coh_flucts}, the averaging time window $T$ should be put similar to the characteristic lifetime of the descending plume. The lifetime of a plume can be estimated from the characteristic turbulent velocity $v$ and the dimension of density stratification. This is much longer than the eddy turnover time $\tau$ estimated from $v$ and mixing length. So, it is expected to be possible to define an averaging time that satisfies the condition (\ref{eq:ave_time_cond}).

	On the other hand, as for the space averaging, we consider
\begin{equation}
	\langle {f} \rangle
	= \frac{1}{\Delta S} \int_{S} f({\textbf{r}};t)\ dS
	\label{eq:space_ave_gen}
\end{equation}
Here $\Delta S$ is the area of the averaging surface, which is spanned by the homogeneous directions. In case the statistical quantity is inhomogeneous in the vertical or $z$ direction, and homogeneous in the horizontal or $x$-$y$ directions, we put $\Delta S = \Delta_x \Delta_y$ and the space averaging is defined with the horizontal averaging as
\begin{equation}
	\langle {f} \rangle(z;t)
	= \frac{1}{\Delta_x \Delta_y} \int_{S} f(x,y,z;t)\ dx dy
	\label{eq:space_ave_hor}
\end{equation}
In this work, the statistical quantities are assumed to be homogeneous in the horizontal surface, and the  horizontal averaging is regarded as a surrogate of the ensemble averaging.

	We adopt the time--space double averaging methodology. In this procedure, we first take the time average of a physical quantity $f$ and denote it as $\overline{f}$. Then we take the space average of the time averaged quantity and denote it as $\langle {\overline{f}} \rangle$. With this double-averaging procedure, a field quantity $f$ is decomposed into
\begin{equation}
	f 
	= \langle {\overline{f}} \rangle
	+ \tilde{f}
	+ f'',
	\label{eq:double_ave_decomp}
\end{equation}
where $\tilde{f}$ is defined by
\begin{equation}
	\tilde{f} = \overline{f} - \langle {\overline{f}} \rangle.
	\label{eq:coh_fluct_def}
\end{equation}
It is often called the dispersion in analogy with the wave decomposition procedure. If we rewrite this in the form of
\begin{equation}
	\overline{f}
	= \langle {\overline{f}} \rangle
	+ \tilde{f},
	\label{eq:space_ave_coh_fluct_def}
\end{equation}
we see that a time averaged quantity $\overline{f}$ is divided into the space averaged part $\langle {\overline{f}} \rangle$ and the deviation from the space average, $\tilde{f}$. Namely, $\tilde{f}$ represents the part of a time-averaged quantity $\overline{f}$ deviating from the space average $\langle {\overline{f}} \rangle$, and $f''$ represents the deviation from the time average $\overline{f}$. On the other hand, from the viewpoint of the space averaging, both $\tilde{f}$ and $f''$ are treated as the fluctuations. The fluctuation in the space averaging procedure, $f'$, is divided into $\tilde{f}$ and $f''$ as
\begin{equation}
	f' = \tilde{f} + f''.
	\label{eq:coh_incoh_fluct_def}
\end{equation}
In other words, $\tilde{f}$ represents the persistent or coherent part of the spatial fluctuations $f'$, while $f''$ is the random or incoherent counterpart.

	Of course, in order to distinguish the coherent fluctuation from the incoherent one, it is required to use much more elaborated mathematical tools. In the wavelet analysis of turbulence, a velocity field is decomposed on the basis of scaling functions and wavelets of different families are often applied. The part of fluctuation is called coherent if it is expressed in terms of such an expansion, and incoherent otherwise \citep{far2003,gol2004}. The notion of coherency does not directly correspond to the usual decomposition between the large and small scales. Coherent motions exist even at small scales, and random motions are observed even at large scales. In this work, however, we do not delve into such detailed formulations of coherency, but denote $\tilde{f}$ as the coherent fluctuation and $f''$ as the incoherent fluctuation in the present time--space double averaging procedure.

\subsection{Relations of time and space averaging}
\label{sec:maths} 

	In the present work, we assume the following relations among the time and space averaging:
\begin{subequations}\label{eq:time-space_ave_rels}
\begin{equation}
	\langle {\langle {f} \rangle} \rangle
	= \langle {f} \rangle,\;\;\;
	\overline{\overline{f}}
	= \overline{f},
	\label{eq:time-space_ave_rey}
\end{equation}
\begin{equation}
	\langle {f'} \rangle = 0,\;\;\;
	\langle {\tilde{f}} \rangle = 0,\;\;\;
	\langle {f''} \rangle = 0,\;\;\;
	\overline{f''} = 0,
	\label{eq:time-space_ave_centered}
\end{equation}
\begin{equation}
	\langle {\overline{f}} \rangle = \langle {f} \rangle,\;\;\;
	\overline{\langle {f} \rangle} = \langle {f} \rangle,
	\label{eq:time-space_ave_order}
\end{equation}
\begin{equation}
	\langle {\langle {f} \rangle \langle {g} \rangle} \rangle
	= \langle {f} \rangle \langle {g} \rangle,\;\;\;
	\overline{\overline{f} \overline{g}}
	= \overline{f} \overline{g},
	\label{eq:time-space_ave_prod1}
\end{equation}
\begin{equation}
	\langle {\overline{f} \overline{g}} \rangle
	= \langle {\overline{f}} \rangle \langle {\overline{g}} \rangle
	= \langle {f} \rangle \langle {g} \rangle,
	\label{eq:time-space_ave_prod2}
\end{equation}
\begin{equation}
	\langle {\tilde{f} g''} \rangle = 0,\;\;\;
	\langle {f'' \tilde{g}} \rangle = 0,
	\label{eq:time-space_ave_co-inco_space}
\end{equation}
\begin{equation}
	\overline{\tilde{f} g}
	= \tilde{f} \overline{g},\;\;\;
	\overline{\tilde{f} g''}
	= 0.
	\label{eq:time-space_ave_co-inco_time}
\end{equation}
\end{subequations}
Among these relations, (\ref{eq:time-space_ave_rey}) is an assumption that the space and time averaging satisfies the Reynolds rule. Since time averaging (\ref{eq:time_ave_short}) contains information of the past time during the averaging period $T$, so strictly speaking, we have $\overline{\overline{f}} \neq \overline{f}$ for a finite $T$. However, we approximate that $\overline{\overline{f}} = \overline{f}$. Relation (\ref{eq:time-space_ave_order}) implies that both the space averaging of the already time-averaged, $\langle {\overline{f}} \rangle$, and the time averaging of the already space-averaged, $\overline{\langle {f} \rangle}$, are equivalent to the space averaged one $\langle {f} \rangle$. This is a consequence of the fact that the time averaging and space averaging are defined independently. Relation (\ref{eq:time-space_ave_co-inco_time}) is derived from (\ref{eq:time-space_ave_rey}) and (\ref{eq:time-space_ave_order}) as
\begin{eqnarray}
	\overline{\tilde{f} g}
	&=& \overline{(\overline{f} - \langle {\overline{f}} \rangle) g}
	= \overline{\overline{f} g} - \overline{\overline{\langle {f} \rangle} g}
	= \overline{f} \overline{g}
	- \overline{\langle {f} \rangle} \overline{g}
	\nonumber\\
	&=& (\overline{f} - \langle {\overline{f}} \rangle) \overline{g}
	= \tilde{f} \overline{g}.
	\label{eq:time-space_ave_co-inco_time_der}
\end{eqnarray}
These relations (\ref{eq:time-space_ave_rels}) are fully utilised in deriving the evolution equations of the turbulent correlations in \S~\ref{sec:eqs_turb_correl}.

\begin{figure}
  \centering
  \includegraphics[width= 0.90 \columnwidth]{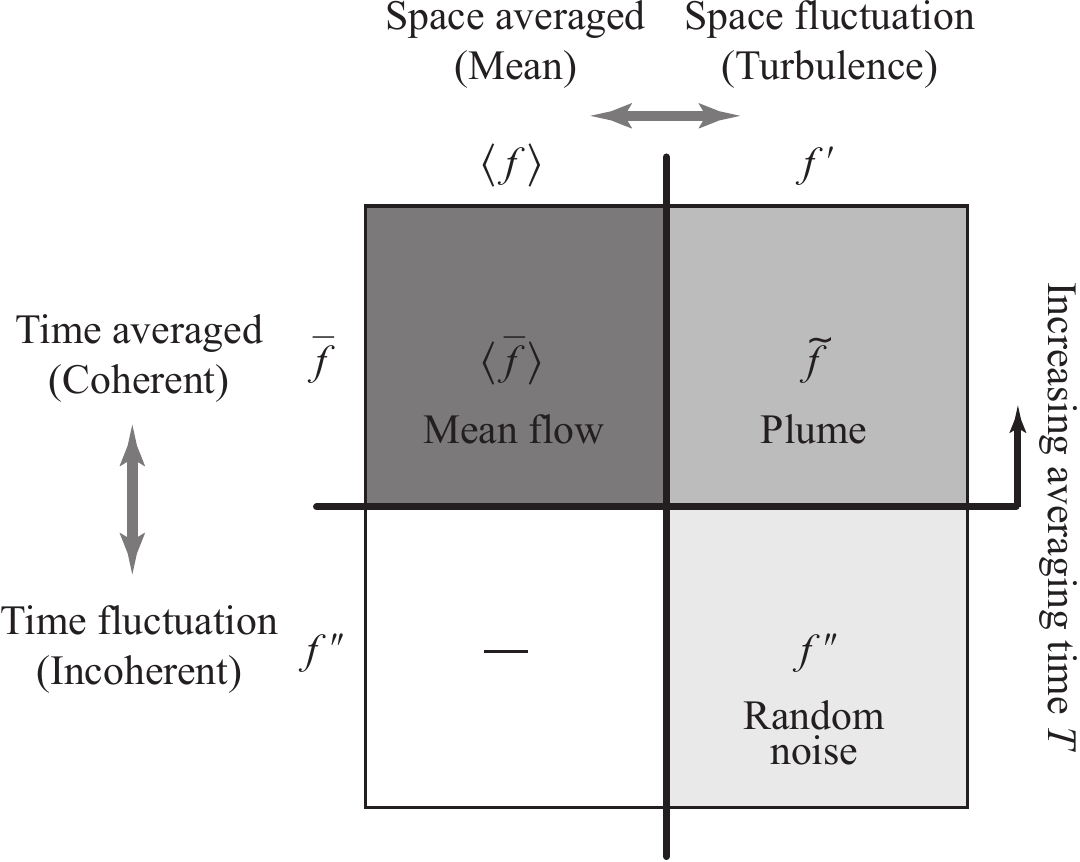}
  \caption{Coherency diagram of the time--space double averaging method. On the basis of the decomposition (\ref{eq:time-space_doub_ave_smry}): $f = \langle {\overline{f}} \rangle + \tilde{f} + f''$, the fluctuation in the space averaging $f'$ is decomposed into the coherent/dispersion and incoherent/random fluctuations, and the time-averaged component is divided into the space-averaged or mean component $\langle {\overline{f}} \rangle$ and the dispersion or deviation from it, $\tilde{f}$.}
    \label{fig:coherency_diagram}
\end{figure}

	We adopt the time--space double averaging procedure. In this framework, a dispersion $\tilde{f}$ belongs to the averaged field $\overline{f} (= \langle {\overline{f}} \rangle + \tilde{f})$ in the time-averaging sense, while it belongs to a fluctuation field $f' (= \tilde{f} + f'')$ and represents the coherent components of fluctuations in the space-averaging sense. In order to see the structure of the turbulence model we adopted in this work, we present a diagram similar to the coherency diagram first introduced by \citet{gol2004}. 
	In the context of the present double-averaging method, this diagram illustrates the relationship among the field quantities decomposed as (\ref{eq:double_ave_decomp}). It shows that the fluctuations are decomposed into the coherent and incoherent fields. It simultaneously exhibits the separation between the space-averaged mean fields and the residual fluctuation fields (Fig.~\ref{fig:coherency_diagram}).

	In the present work, a time-averaged quantity $\overline{f}$ is divided into its space-averaged part $\langle {\overline{f}} \rangle$ and its deviation or dispersion part $\tilde{f} (= \overline{f} - \langle {\overline{f}} \rangle)$. The space averaging of the incoherent fluctuation vanishes by definition. This can be seen from (\ref{eq:time-space_ave_rels}) as $\langle {f''} \rangle = \langle {f - \overline{f}} \rangle = \langle {f} \rangle - \langle {\overline{f}} \rangle = 0$. The criterion for separating the coherent component from incoherent one depends on the averaging time $T$ in (\ref{eq:time_ave_short}). If we set the averaging time $T$ sufficiently large (e.g., much larger than the lifetime of the characteristic coherent fluctuation such as a plume), the coherent motions are smeared out and the portion of the coherent component is reduced. This situation is not suitable for properly describing the effects of plume motions, which is one of the essential ingredients of convective flow.

	In summary, with the time--space averaging procedure, a field quantity $f$ is decomposed as
\begin{equation}
    f 
	= \overbrace{
		\langle {\overline{f}} \rangle
		+ \underbrace{\tilde{f}}_
	{
	\begin{array}{c}
		\overline{f} - \langle {\overline{f}} \rangle\\
		\mbox{coherent}\\
		\mbox{fluctuation}
	\end{array}
	}
	}^{\overline{f}}
	+ \underbrace{f''}_{
	\begin{array}{c}
	f - \overline{f}\\
	\mbox{incoherent}\\
	\mbox{fluctuation}
	\end{array}}.
	\label{eq:time-space_doub_ave_smry}
\end{equation}

\section{Evolution equations of turbulence correlations}\label{sec:eqs_turb_correl}
	The evolution equations of the coherent and incoherent components of the Reynolds stress are given by
\begin{Eqnarray}
	\left( {
		\frac{\partial}{\partial t}
		+ \langle {u} \rangle ^\ell 
		\frac{\partial} {\partial x^\ell}
	} \right) \langle {\tilde{u}^i \tilde{u}^j} \rangle
	&=& \tilde{P}^{ij}
	+ \tilde{\Pi}^{ij}
	- \tilde{\varepsilon}^{ij}
	+ \frac{\partial \tilde{T}^{ij\ell}}{\partial x^\ell}
	\nonumber\\
	&&\hspace{-50pt}+ \left\langle {
		\widetilde{u''{}^\ell u''{}^j}
		\frac{\partial \tilde{u}^i}{\partial x^\ell}
	} \right\rangle
	+ \left\langle {
		\widetilde{u''{}^\ell u''{}^i}
		\frac{\partial \tilde{u}^j}{\partial x^\ell}
	} \right\rangle,
	\label{eq:coh_rey_strss_eq_gen}
\end{Eqnarray}
\begin{Eqnarray}
	\left( {
		\frac{\partial}{\partial t}
		+ \langle {u} \rangle ^\ell 
		\frac{\partial} {\partial x^\ell}
	} \right) \langle {u''{}^i u''{}^j} \rangle
	&=& P''{}^{ij}
	+ \Pi''{}^{ij}
	- \varepsilon''{}^{ij}
	+ \frac{\partial T''{}^{ij\ell}}{\partial x^\ell}
	\nonumber\\
	&&\hspace{-50pt}- \left\langle {
		\widetilde{u''{}^\ell u''{}^j}
		\frac{\partial \tilde{u}^i}{\partial x^\ell}
	} \right\rangle
	- \left\langle {
		\widetilde{u''{}^\ell u''{}^i}
		\frac{\partial \tilde{u}^j}{\partial x^\ell}
	} \right\rangle.
	\label{eq:incoh_rey_strss_eq_gen}
\end{Eqnarray}
Here, $\tilde{P}^{ij}$, $\Pi^{ij}$, and $\partial \tilde{T}^{ij\ell}/\partial x^\ell$ are the production, re-distribution, dissipation, and transport rates of the coherent Reynolds stress $\langle {\tilde{u}^i \tilde{u}^j} \rangle$, respectively, and $P''{}^{ij}$, $\Pi''{}^{ij}$, $\varepsilon''{}^{ij}$, and $\partial T''{}^{ij\ell}/\partial x^\ell$ are the counterparts for the incoherent Reynolds stress $\langle {u''{}^i u''{}^j} \rangle$. They are in a similar form as the counterparts in the equation of the usual Reynolds stress $\langle {u'{}^i u'{}^j} \rangle$. The detailed expressions of these terms are given in Appendix~\ref{sec:append_B}. In contrast, the final two terms in (\ref{eq:coh_rey_strss_eq_gen}) and (\ref{eq:incoh_rey_strss_eq_gen}) are ones newly appeared in the time--space averaging procedure. Firstly, in the presence of inhomogeneous dispersion velocity, $\nabla \tilde{\bf{u}}$, coupled with the dispersion of the Reynolds stress $\widetilde{{\bf{u}}'' {\bf{u}}''}$ [defined by (\ref{eq:disp_incoh_strss})], the coherent and incoherent Reynolds stresses, $\langle {\tilde{\bf{u}} \tilde{\bf{u}}} \rangle$ and $\langle {{\bf{u}}'' {\bf{u}}''} \rangle$, can be produced or reduced. This is an interesting point. The presence of a localised plume structure itself interacts with the dynamics and statistics of the coherent and incoherent fluctuations. Secondly, the final two terms in (\ref{eq:coh_rey_strss_eq_gen}):
\begin{equation}
	\tilde{\cal{P}}^{ij}
	\equiv + \left\langle {
		\widetilde{u''{}^\ell u''{}^j}
		\frac{\partial \tilde{u}^i}{\partial x^\ell}
	} \right\rangle
	+ \left\langle {
		\widetilde{u''{}^\ell u''{}^i}
		\frac{\partial \tilde{u}^j}{\partial x^\ell}
	} \right\rangle,
	\label{prod_disp}
\end{equation}
and the final two terms in (\ref{eq:incoh_rey_strss_eq_gen}): 
	\begin{equation}
	{\cal{P}}''{}^{ij}
	\equiv - \left\langle {
		\widetilde{u''{}^\ell u''{}^j}
		\frac{\partial \tilde{u}^i}{\partial x^\ell}
	} \right\rangle
	- \left\langle {
		\widetilde{u''{}^\ell u''{}^i}
		\frac{\partial \tilde{u}^j}{\partial x^\ell}
	} \right\rangle
	= - \tilde{\cal{P}}^{ij}
	\label{prod_random}
\end{equation}
are exactly the same but with the opposite sign. This means that neither $\tilde{\mbox{\boldmath$\cal{P}$}} = \{\tilde{\cal{P}}^{ij}\}$ nor $\mbox{\boldmath$\cal{P}$}'' = \{ {\cal{P}}''{}^{ij} \}$ will contribute to the evolution of the total Reynolds stress $\langle {{\bf{u}}' {\bf{u}}'} \rangle$, but they will contribute to the transfer between the coherent and incoherent components, $\langle {\tilde{\bf{u}} \tilde{\bf{u}}} \rangle$ and $\langle {{\bf{u}}'' {\bf{u}}''} \rangle$. In the case of a positive (or negative) $\tilde{\mbox{\boldmath$\cal{P}$}}$, the coherent component of the Reynolds stress, $\langle {\tilde{\bf{u}} \tilde{\bf{u}}} \rangle$, is increased (or decreased) while the incoherent counterpart $\langle {{\bf{u}}'' {\bf{u}}''} \rangle$ is decreased (or increased).
	
	Taking the contraction of $i$ and $j$ in (\ref{eq:coh_rey_strss_eq_gen}) and (\ref{eq:incoh_rey_strss_eq_gen}), we obtain the evolution equations of the coherent and incoherent components of the turbulent fluctuation energy as
\begin{equation}
	\left( {
		\frac{\partial}{\partial t}
		+ \langle {u} \rangle ^\ell \frac{\partial} {\partial x^\ell}
	} \right) \left\langle {\frac{1}{2}\tilde{\bf{u}}^2} \right\rangle
	= \tilde{P}
	- \tilde{\varepsilon}
	+ \nabla \cdot \tilde{\bf{T}}
	+ \left\langle {
		\widetilde{u''{}^\ell u''{}^i}
		\frac{\partial \tilde{u}^i}{\partial x^\ell}
	} \right\rangle,
	\label{eq:coh_en_eq_gen}
\end{equation}
\begin{equation}
	\left( {
		\frac{\partial}{\partial t}
		+ \langle {u} \rangle ^\ell \frac{\partial} {\partial x^\ell}
	} \right) \left\langle {\frac{1}{2}{\bf{u}}''{}^2} \right\rangle
	= P''
	- \varepsilon''
	+ \nabla \cdot {\bf{T}}''
	- \left\langle {
		\widetilde{u''{}^\ell u''{}^i}
		\frac{\partial \tilde{u}^i}{\partial x^\ell}
	} \right\rangle,
	\label{eq:incoh_en_eq_gen}
\end{equation}
respectively. Here, $\tilde{P}$, $\tilde{\varepsilon}$, $\nabla \cdot \tilde{\bf{T}}$ are the production, dissipation, and transport rates of the coherent fluctuation energy, and $P''$, $\varepsilon''$, and $\nabla \cdot {\bf{T}}''$ are the incoherent counterparts. All these terms are in the same form as the equation of the total fluctuation energy $\langle {{\bf{u}}'{}^2} \rangle/2$. Their detailed expressions are given in Appendix~\ref{sec:append_B}.
	
	The final terms in (\ref{eq:coh_en_eq_gen}) and (\ref{eq:incoh_en_eq_gen}) are originated from the double-averaging procedure. The terms related to the coherent fluctuation shear, $\nabla \tilde{\bf{u}}$: 
\begin{equation}
	P_{\tilde{K}}
	\equiv \left\langle {
    	\widetilde{u''{}^\ell u''{}^i}
    	\frac{\partial \tilde{u}^i}{\partial x^\ell}
	} \right\rangle
	= - P_{K''}
	\label{eq:prod_coh_incoh_en}
\end{equation}
represents the production of the coherent energy $P_{\tilde{K}}$ in (\ref{eq:coh_en_eq_gen}) and that of incoherent energy $- P_{K''}$ in (\ref{eq:incoh_en_eq_gen}) due to the shear of the coherent fluctuation. The energy transfer between the coherent- and incoherent-fluctuation energies, $\langle {\tilde{\bf{u}}^2} \rangle/2$ and $\langle {{\bf{u}}''{}^2} \rangle/2$, is attributed to these production rates $P_{\tilde{K}}$ and $P_{K''}$.

	The internal transfer term mediated by the dispersion stress $\widetilde{{\bf{u}}'' {\bf{u}}''}$ always show up when we decompose a field quantity as in (\ref{eq:double_ave_decomp}). Exploring the properties of transfer production terms $P_{\tilde{K}}$ and $P_{K''}$, we understand what mechanisms determine the evolution of the coherent fluctuations. For the same shear structure of the coherent velocity, variations of the coherent and incoherent components of energy is opposite to each other. A decrease of incoherent or random fluctuation energy leads to an increase of the coherent fluctuation energy, and vice versa. As will be seen in the following sections, this energy transfer between the coherent and incoherent fluctuations coupled with the non-equilibrium effect associated with plume motions is expected to contribute to the enhancement of turbulent transport beyond the usual local gradient-diffusion approximation models.

\section{Modelling convective turbulence}\label{sec:model_conv}
\subsection{Non-equilibrium effect}\label{sec:non-equiv}
In the convective turbulent flows, not only the turbulent transport due to the usual eddy or random/incoherent fluctuation but also the transport due to the structural/coherent fluctuation plays a key role. In the presence of plume structures, effective transports of the mass, momentum, and energy are greatly enhanced, compared to the case without the plume structures or to the case with the structures rapidly disappearing. Therefore, how to incorporate the effects of coherent fluctuations such as plumes is of primary importance in modelling the convective turbulent flows.

In the $K-\varepsilon$ model, one of the most widely used Reynolds-averaged turbulence models, the eddy viscosity is expressed as
\begin{equation}
	\nu_{\textrm{TE}}
	= C_\nu \frac{K^2}{\varepsilon},
	\label{eq:eddy_visc_model}
\end{equation}
where $C_\nu$ is the model constant, $K$ is the turbulent kinetic energy defined by (\ref{eq:K_def}), and $\varepsilon$ is its dissipation rate defined by (\ref{eq:eps_def}). They are the two one-point turbulent statistical quantities chosen to describe the dynamic and statistical properties of turbulent fields. The eddy-viscosity model (\ref{eq:eddy_visc_model}) is very simple and useful, but is known to have several drawbacks. One is the overestimate of $K$ and $\varepsilon$ when applied to the homogeneous shear turbulence. In order to alleviate such a drawback, various ways for improving the model  (\ref{eq:eddy_visc_model}) have been proposed. The non-equilibrium effect is one of such methods and is expected to be valid in deriving the turbulent transport that deviates from the simple eddy-viscosity representation \citep{yos1993}. 

	In order to see the background of the non-equilibrium model of turbulent transport, here we briefly show how to derive the non-equilibrium effect from the analytical statistical theory of inhomogeneous turbulence. With the aid of the multiple-scale renormalisation perturbation expansion \citep{yos1984,yok2020}, the turbulent energy with the non-equilibrium effect implemented is expressed as
\begin{equation}
	K 
	= C_{K1} \varepsilon^{2/3} \ell_{\rm{C}}^{2/3}
	- C_{K2} \varepsilon^{-3/2} \ell_{\rm{C}}^{4/3} 
		\frac{D\varepsilon}{Dt}
	- C_{K3} \varepsilon^{1/3} \ell_{\rm{C}}^{1/3}
		\frac{D\ell_{\rm{C}}}{Dt},
	\label{eq:K_non-equil_exp}
\end{equation}
where $\ell_{\rm{C}}$ is the scale of the largest eddy, $C_{Kn}$ ($n=1-3$) are the model constants, and the Lagrangian derivative along the mean velocity ${\bf{U}} (= \langle {\bf{u}} \rangle)$ is defined as
\begin{equation}
	\frac{D}{Dt} 
	= \frac{\partial}{\partial t}
	+ {\bf{U}} \cdot \nabla.
	\label{eq:lag_der_mean_vel}
\end{equation}

We solve (\ref{eq:K_non-equil_exp}) by iteration. The lowest-order expression is
\begin{equation}
	K = C_{K1} \varepsilon^{2/3} \ell_{\rm{C}}^{2/3},
	\label{eq:K_lowest_exp}
\end{equation}
leading to $\ell_{\rm{C}}$ as
\begin{equation}
	\ell_{\textrm{C}}
	= C_{\ell 1} K^{3/2} \varepsilon^{-1}
	\label{eq:ell_lowest_exp}
\end{equation}
($C_{\ell 1}$: model constant). This expression, obtained without recoursing to the non-equilibrium part, corresponds to the simplest eddy-viscosity expression. Substituting (\ref{eq:ell_lowest_exp}) into (\ref{eq:K_non-equil_exp}), and using the iteration, we have
\begin{equation}
	\ell_{\rm{C}}
	= C_{\ell 1} K^{3/2} \varepsilon^{-1}
	+ C_{\ell 2} K^{3/2} \varepsilon^{-2} \frac{DK}{Dt}
	- C_{\ell 3} K^{5/2} \varepsilon^{-3} \frac{D\varepsilon}{Dt}
	\label{eq:ell_non-equib_exp}
\end{equation}
($C_{\ell 2}$ and $C_{\ell 3}$: model constants). The second and third terms in (\ref{eq:ell_non-equib_exp}) represent the non-equilibrium effect. Equation~(\ref{eq:ell_non-equib_exp}) can be approximated as
\begin{equation}
	\ell_{\rm{C}}
	= \ell_{\rm{E}} \left( {
		1 - C'_{\rm{N}} \frac{1}{K} 
		\frac{D}{Dt} \frac{K^2}{\varepsilon}
	} \right),
	\label{eq:ell_non-equib_Dt_exp}
\end{equation}
where $\ell_{\rm{E}}$ is the equilibrium length scale defined by
\begin{equation}
	\ell_{\textrm{E}} = K^{3/2}/\varepsilon
	\label{eq:ell_equib}
\end{equation}
($C'_{\rm{N}}$: model constant). It follows from (\ref{eq:ell_non-equib_Dt_exp}) that the time scale of turbulence is expressed as
\begin{equation}
	\tau_{\textrm{NE}}
	= \frac{\ell_{\textrm{NE}}}{K^{1/2}}
	\simeq \tau_{\textrm{E}} 
		\left( {
			1 - C'_{\textrm{N}} \Gamma
		} \right),
	\label{eq:tau_non-equib}
\end{equation}
where the non-equilibrium correction factor $\Gamma$ is defined by
\begin{equation}
	\Gamma = \frac{1}{K} \frac{D}{Dt} \frac{K^2}{\varepsilon},
	\label{eq:Gamma_def}
\end{equation}
which can be either positive or negative. Here, $\tau_{\rm{E}}$ is the equilibrium time scale defined by
\begin{equation}
	\tau_{\textrm{E}}
	= \frac{K}{\varepsilon}.
	\label{eq:tau_equib}
\end{equation}
As we see from (\ref{eq:ell_non-equib_Dt_exp}) and (\ref{eq:tau_non-equib}), the non-equilibrium effect can be represented by the variation of the turbulent energy $K$ and its dissipation rate $\varepsilon$ along the fluid motion. From (\ref{eq:tau_non-equib}), the turbulent viscosity is expressed as
\begin{eqnarray}
	\everymath{\displaystyle}\nu_{\textrm{T}}
	= \left\{ {
	\begin{array}{lll}
		\nu_{\textrm{TE}} \left( {
			1 + C_{\textrm{N}} \Gamma
		} \right)^{-1}
	&\mbox{for}
	&\Gamma > 0,
	\\
	\rule{0.ex}{6.ex}
	\nu_{\textrm{TE}} \left( {
		1 - C_{\textrm{N}} \Gamma
	} \right)
	&\mbox{for}
	& \Gamma < 0,
	\end{array}
	} \right.
	\label{eq:nu_T_non-equib_exp}
\end{eqnarray}
where $\nu_{\rm{TE}}$ is the equilibrium eddy viscosity without the non-equilibrium effect related to $\Gamma$ ($C_{\rm{N}}$: model constant).

	Expression~(\ref{eq:nu_T_non-equib_exp}) implies that in case that  the turbulent energy $K$ and its dissipation rate $\varepsilon$ vary along the mean flow, namely in the case of non-equilibrium turbulence, the effective turbulent viscosity $\nu_{\rm{T}}$ may deviate from its equilibrium value. Equation~(\ref{eq:nu_T_non-equib_exp}) shows that the effective viscosity is suppressed if $\Gamma > 0$ or equivalently $D(K^2/\varepsilon)/Dt = D(K\tau)/Dt > 0$ with the time scale $\tau = K/\varepsilon$, and that $\nu_{\rm{T}}$ is enhanced if $\Gamma < 0$ or equivalently $D(K^2 / \varepsilon)/Dt = D(K\tau)/Dt < 0$.

	Actually, it was established by recent experiments with single-phase fluid jets and with two-phase fluid buoyant bubble plumes and negatively buoyant particle plumes that the turbulent kinetic energy $K$ and its dissipation rate $\varepsilon$ vary along the plume and jet stream direction \citep{bor2020,cha2017,lai2019a,lai2019b}. For instance, in the round jet experiment, it was reported that the ratio of the axial fluctuation intensity $\langle {u'{}^2} \rangle$ to the root of the dissipation rate, $\sqrt{\varepsilon}$, showed a decreasing tendency with the axial or streamwise direction \citep{lai2019a}. In this case, since the non-equilibrium contribution $\Gamma$ in (\ref{eq:Gamma_def}) is negative, $D(K^2/\varepsilon)/Dt < 0$, it is anticipated from (\ref{eq:nu_T_non-equib_exp}) that the turbulent viscosity will be effectively enhanced. This matches the requirement for improving the current turbulence model by adjusting the model constants. For another example, in the buoyant bubble plume experiment, it was reported that the turbulent energy is increased along the ascending bubble motion. This tendency is much more dominant in the adjustment region, where the non-equilibrium property is much more prominent, than in the asymptotic region \citep{lai2019b}. In another experiment, the turbulent energy is reported to be enhanced in the direction of the negatively buoyant particle plume, while the energy dissipation does not show significant changes \citep{bor2020}. In these two cases of the buoyant bubble plume and negatively buoyant particle plume, $D(K^2/\varepsilon)/Dt$ is positive, so the turbulent viscosity is expected to be reduced by the non-equilibrium effect. As we see from these jet and plume experiments, the inhomogeneities of the turbulent kinetic energy and its dissipation rate along the flow direction may lead to a significant alteration of the turbulent transport coefficients due to the non-equilibrium effect \citep{yok2022}.

\subsection{Modelling plume with double averaging}\label{sec:plume_double_ave}
For the sake of simplicity of argument, we further assume that the non-equilibrium property along a flow for the energy dissipation rate $\varepsilon$ is much smaller than the counterpart on the turbulent energy $K (= \langle {{\bf{u}}}'{}^2 \rangle/2)$ in the sense
\begin{equation}
	\frac{1}{\varepsilon} \frac{D\varepsilon}{Dt}
	\ll \frac{1}{K} \frac{DK}{Dt}.
	\label{eq:Deps_ll_DK_cond}
\end{equation}
Under this condition, the non-equilibrium eddy viscosity (\ref{eq:nu_T_non-equib_exp}) is approximately expressed as
\begin{equation}
	\nu_{\rm{NE}} 
	= \nu_{\rm{E}} \left( {
		1 - C_{\rm{N}} \frac{1}{\varepsilon} \frac{DK}{Dt}
	} \right)
	= \nu_{\rm{E}} \left( {
		1 - C_{\rm{N}} \frac{\tau}{K} \frac{DK}{Dt}
	} \right),
	\label{eq:nu_T_simpl}
\end{equation}
where $\tau$ is the eddy turn-over time, and $C_{\rm{N}}$ the model constant. Inequality (\ref{eq:Deps_ll_DK_cond}) does not necessarily meet the flow conditions in all the real turbulence cases. However, we adopt (\ref{eq:nu_T_simpl}) as a starting point, since the estimate of dissipation rate in practical flows is often much harder than that of the turbulent energy.

	With the analogy of the non-equilibrium effect on the Reynolds stress in the ensemble- or space-averaging procedure, (\ref{eq:nu_T_simpl}), the turbulent mass flux $\overline{\rho'' {\bf{u}}''}$ in the time-averaging procedure may be modelled as
\begin{equation}
	\overline{\rho'' {\textbf{u}}''}
	= - \kappa''_{\textrm{E}} \left( {
		1 
		- C'' \tau'' 
		\frac{1}{k''} 
		\frac{\overline{D}k''}{\overline{D}t}
	} \right) \nabla \overline{\rho},
	\label{eq:rand_mass_flux}
\end{equation}
where $\kappa''_{\rm{E}}$ is the equilibrium effective diffusivity due to the random or incoherent fluctuations, $C''$ the model constant, $\tau''$ the time scale of the incoherent fluctuation, $\overline{D}/\overline{D} t (= \partial/\partial t + \overline{\bf{u}} \cdot \nabla)$ the Lagrange or material derivative along the time-averaged velocity $\overline{\bf{u}}$, and $k''$ the time-averaged turbulent energy of the incoherent fluctuation defined by
\begin{equation}
	k'' = \overline{{\textbf{u}}''{}^2}/2.
	\label{eq:rand_en_def}
\end{equation}
By a similar analogy, the dispersion turbulent mass flux $\overline{\tilde{\rho} \tilde{\textbf{u}}}$ may be modelled as
\begin{equation}
	\overline{\tilde{\rho} \tilde{\bf{u}}}
	= - \tilde{\kappa}_{\rm{E}} \left( {
		1 
		- \tilde{C} \tilde{\tau} 
    		\frac{1}{\tilde{k}} 
    		\frac{\overline{D}\tilde{k}}{\overline{D}t}
	} \right) \nabla \overline{\rho},
	\label{eq:disp_mass_flux}
\end{equation}
In relation to (\ref{eq:double_ave_decomp}) and (\ref{eq:space_ave_coh_fluct_def}), the total time-averaged turbulent energy is defined by
\begin{equation}
	k = \overline{{\textbf{u}}'{}^2}/2.
	\label{eq:time_ave_turb_en_def}
\end{equation}
If we assume that the coherent and incoherent fluctuations have no correlation in time: $\overline{\tilde{f} g''} = 0$ [see (\ref{eq:time-space_ave_co-inco_time})], the total turbulent energy and the coherent and incoherent fluctuation energies are related to each other as\footnote{The assumption of no correlation between the coherent and incoherent fluctuations in time is equivalent to the present formulation (\ref{eq:double_ave_decomp}) with the decomposition of a field quantity $f'$ into $ \tilde{f}$ and $f''$ (\ref{eq:coh_incoh_fluct_def}). In this sense, assumption of no-correlation between $\tilde{f}$ and $g''$ (or $f''$) is just the restatement of the adoption of the decomposition (\ref{eq:coh_incoh_fluct_def}). Note that this does not deny the interaction between the coherent and incoherent fluctuations at all. As  (\ref{eq:prod_coh_incoh_en}) shows, the production of the incoherent fluctuation energy results from the sink of the coherent fluctuation energy.}
\begin{subequations}\label{eq:time_ave_en_rel}
\begin{equation}
	k = \tilde{k} + k'',
	\label{eq:k_disp_+_k_rand}
\end{equation}
or equivalently, 
\begin{equation}
	\overline{{\bf{u}}'{}^2} 
	= \overline{\tilde{\bf{u}}^2} 
	+ \overline{{\bf{u}}''{}^2}.
	\label{eq:en_disp_+_en_rand}
\end{equation}
\end{subequations}
	In the time--space double averaging procedure, the turbulent mass flux $\langle {\rho' {\bf{u}}'} \rangle (= \left\langle {\overline{\rho' {\textbf{u}}'}} \right\rangle)$ is divided into the contribution from the coherent fluctuations and the incoherent ones. Assuming that each contribution to the turbulent mass flux is expressed as (\ref{eq:rand_mass_flux}) and (\ref{eq:disp_mass_flux}), we can write the turbulent mass flux as
\begin{Eqnarray}
	\left\langle {\overline{\rho' {\textbf{u}}'}} \right\rangle
	&=& \left\langle {
		\overline{\tilde{\rho} \tilde{\textbf{u}}}
	} \right\rangle
	+ \left\langle {\overline{\rho'' \textbf{u}''}} \right\rangle
	\nonumber\\
	&=& - \left\langle {
		\tilde{\kappa}_{\textrm{E}} \left( {
			1 
			- \tilde{C} \tilde{\tau} 
			\frac{1}{\overline{\tilde{\textbf{u}}^2}} 
			\frac{\overline{D}\overline{\tilde{\textbf{u}}^2}}{\overline{D}t}
		} \right) \nabla \overline{\rho}
	} \right\rangle
	\nonumber\\
	&&- \left\langle {
		\kappa''_{\textrm{E}} \left( {
		1 
		- C'' \tau'' 
			\frac{1}{\overline{{\textbf{u}}''{}^2}} 
			\frac{\overline{D} \overline{{\textbf{u}}''{}^2}}
			{\overline{D}t}
		} \right) \nabla \overline{\rho}
	} \right\rangle.
	\label{eq:turb_mass_flux}
\end{Eqnarray}
The first or unity-related part in the parentheses of each of these two terms gives the usual eddy diffusivity contribution to the turbulent mass flux. They represent the equilibrium effect in the turbulent mass flux and are written as
\begin{equation}
	\left\langle {\rho' {\bf{u}}'} \right\rangle_{\rm{E}}
	= - \kappa_{\rm{E}} \nabla \left\langle {\overline{\rho}} \right\rangle,
	\label{eq:turb_mass_flux_equib}
\end{equation}
where $\kappa_{\rm{E}} = \tilde{\kappa}_{\rm{E}} + \kappa''_{\rm{E}}$ is the equilibrium turbulent eddy diffusivity. The rest parts of two terms of Eq. (\ref{eq:turb_mass_flux}) or the $\overline{D}/\overline{D}t$-related parts in the parentheses represent the non-equilibrium effect. They are written as
\begin{eqnarray}
	\lefteqn{
	\left\langle {\overline{\rho' {\textbf{u}}'}} \right\rangle_{\rm{N}}
	= + \left\langle {
		\tilde{\kappa}_{\rm{E}}
		\tilde{C} \tilde{\tau} 
			\frac{1}{\overline{\tilde{\bf{u}}^2}} 
			\frac{\overline{D}\overline{\tilde{\bf{u}}^2}}{\overline{D}t} 
				\nabla \overline{\rho}
	} \right\rangle
	+ \left\langle {
		\kappa''_{\rm{E}}
			C'' \tau''
			\frac{1}{\overline{{\bf{u}}''{}^2}} 
			\frac{\overline{D} \overline{{\bf{u}}''{}^2}}
				{\overline{D}t} \nabla \overline{\rho}
	} \right\rangle
	}\nonumber\\
	&& \hspace{-35pt} = \left\langle {
		\tilde{\kappa}_{\rm{E}}
		\tilde{C} \frac{\tilde{\tau}}{\overline{\tilde{\bf{u}}^2}} 
			\left[ {
				(\tilde{\bf{u}} \cdot \nabla) \overline{\tilde{\bf{u}}^2}
			} \right]
				\nabla \overline{\rho}
	} \right\rangle
	+ \left\langle {
		\kappa''_{\rm{E}}
			C'' \frac{\tau''}{\overline{{\bf{u}}''{}^2}} 
			\left[ {
				(\tilde{\bf{u}} \cdot \nabla) \overline{{\bf{u}}''{}^2}
			} \right]]
				\nabla \overline{\rho}
	} \right\rangle,
	\label{eq:turb_mass_flux_nonequil}
\end{eqnarray}
where the time derivative part $\partial/\partial t$ of the Lagrangian derivative was dropped since the time derivatives of the time-averaged quantities are much smaller than the advective derivative part $\tilde{\bf{u}} \cdot \nabla$.\footnote{The effects of the time derivatives are different between the fast and slow time scales. To see this we introduce two-scale time variables for time $t$, the fast and slow variables are given by $\tau (= t)$ and $T (= \delta t)$, respectively, where $\delta (\ll 1)$ is a scale parameter. A time-averaged quantity $\overline{f}$ depends only on the slow variable $T$ as $\overline{f} = \overline{f}(T)$. Under this two-time analysis, the time derivative is written as $\partial / \partial t = \partial / \partial \tau + \delta \partial / \partial T$. Then the time derivative of the time-averaged quantity is expressed as $\partial \overline{f} / \partial t = (\partial / \partial \tau + \delta \partial \overline{f} / \partial T) \overline{f} = \delta \partial \overline{f} / \partial T$. Because of the smallness of $\delta$, the time derivative of the time-averaged quantity is eventually very small. On the other hand, the space derivative of the time-averaged quantity is not small since the scales associated with the space derivative is much larger than the counterparts of the time averaging.} According to the relation (\ref{eq:time-space_ave_rels}), the double-averaging procedure obeys
\begin{equation}
	\langle {\overline{f} \overline{g}} \rangle
	= \langle {
		(\langle {\overline{f}} \rangle + \tilde{f})
		(\langle {\overline{g}} \rangle + \tilde{g})
	} \rangle
	= \langle {\overline{f}} \rangle \langle {\overline{g}} \rangle
	+ \langle {\tilde{f} \tilde{g}} \rangle,
	\label{eq:time_ave_fg_prod}
\end{equation}
\begin{eqnarray}
	\lefteqn{
	\langle {\overline{f} \overline{g} \overline{h}} \rangle
	= \langle {
		(\langle {\overline{f}} \rangle + \tilde{f})
		(\langle {\overline{g}} \rangle + \tilde{g})
		(\langle {\overline{h}} \rangle + \tilde{h})
	} \rangle
	}\nonumber\\
	&&\hspace{-15pt}= \langle {\overline{f}} \rangle 
		\langle {\overline{g}} \rangle
		\langle {\overline{h}} \rangle
		+ \langle {\overline{f}} \rangle 
			\langle {\tilde{g} \tilde{h}} \rangle
		+ \langle {\overline{g}} \rangle 
			\langle {\tilde{f} \tilde{h}} \rangle
		+ \langle {\overline{h}} \rangle 
			\langle {\tilde{f} \tilde{g}} \rangle
		+ \langle {\tilde{f} \tilde{g} \tilde{h}} \rangle.
	\label{eq:time_ave_fgh_prod}
\end{eqnarray}
If the space average of the time average is small as $\langle {\overline{f}} \rangle \simeq 0$, we have $\overline{f} = \langle {\overline{f}} \rangle + \tilde{f} \simeq \tilde{f}$. In this case, (\ref{eq:time_ave_fg_prod}) and (\ref{eq:time_ave_fgh_prod}) are reduced to
\begin{equation}
	\langle {\overline{f} \overline{g}} \rangle
	= \langle {\tilde{f} \overline{g}} \rangle
	= \langle {\tilde{f} \tilde{g}} \rangle,
	\label{eq:fg_prod_simpl}
\end{equation}
\begin{equation}
	\langle {\overline{f} \overline{g} \overline{h}} \rangle
	= \langle {\tilde{f} \overline{g} \overline{h}} \rangle
	= \langle {\overline{g}} \rangle 
		\langle {\tilde{f} \tilde{h}} \rangle
	+ \langle {\overline{h}} \rangle 
		\langle {\tilde{f} \tilde{g}} \rangle
	+ \langle {\tilde{f} \tilde{g} \tilde{h}} \rangle.
	\label{eq:fgh_prod_simpl}
\end{equation}
For a convective flow in a closed symmetric system, the space-averaged velocity is expected to be small ($\langle {\overline{\bf{u}}} \rangle \simeq 0$). In this case, the time average of ${\bf{u}}$ is represented by the dispersion part of the velocity as 
\begin{equation}
	\overline{\bf{u}} 
	= \langle {\overline{\bf{u}}} \rangle + \tilde{{\bf{u}}} 
	\simeq \tilde{{\bf{u}}}.
	\label{time_ave_dis_vel}
\end{equation}	
With (\ref{eq:fg_prod_simpl}) and (\ref{eq:fgh_prod_simpl}), the non-equilibrium correction to the eddy diffusivity is written as
\begin{eqnarray}
	\lefteqn{
	\kappa_{\textrm{N}}
	= + \tilde{C} \tilde{\kappa}_{\textrm{E}} \left\langle { 
    	\frac{\tilde{\tau}}{\overline{\tilde{\textbf{u}}^2}} 
    	\frac{\overline{D}\overline{\tilde{\textbf{u}}^2}}{\overline{D}t}
	} \right\rangle
	+ C'' \kappa''_{\textrm{E}} \left\langle { 
		\frac{\tau''}{\overline{{\textbf{u}}''{}^2}} 
		\frac{\overline{D} \overline{{\textbf{u}}''{}^2}}{\overline{D}t}
		} \right\rangle
	}\nonumber\\
	&&\hspace{-20pt} = \tilde{C} \frac{\tilde{\tau}}{\left\langle {
			\overline{\tilde{\textbf{u}}^2}
		} \right\rangle}
	\left\langle {
		\left( {
			\tilde{\textbf{u}} \cdot \nabla
		} \right) \overline{\tilde{{\textbf{u}}}^2}
	} \right\rangle
	+ \tilde{C} \left\langle {
		\frac{\tilde{\tau}}{\overline{\tilde{\textbf{u}}^2}}
		\left( {
			\tilde{\textbf{u}} \cdot \nabla
		} \right)
	} \right\rangle \left\langle {
		\overline{\tilde{\textbf{u}}^2}
	} \right\rangle
	\nonumber\\
	&&\hspace{-20pt} + C'' \frac{\tau''}{\left\langle {
		\overline{{\textbf{u}''}^2}
		} \right\rangle}
	\left\langle {
		\left( {
		\tilde{\textbf{u}} \cdot \nabla
		} \right) \overline{{\textbf{u}}''{}^2}
	} \right\rangle
	+ C'' \left\langle {
		\frac{\tau''}{\overline{{\textbf{u}}''{}^2}}
			\left( {
				\tilde{\textbf{u}} \cdot \nabla
			} \right)
	} \right\rangle \left\langle {\overline{\textbf{u}''{}^2}} \right\rangle
	\nonumber\\
	&&\hspace{-20pt} \simeq \tilde{C} \frac{\tilde{\tau}}
		{\left\langle {\tilde{\textbf{u}}^2} \right\rangle}
		\left\langle {
			\left( {
				\tilde{\textbf{u}} \cdot \nabla
			} \right) \overline{\tilde{{\textbf{u}}}^2}
		} \right\rangle
	+ C'' \frac{\tau''}{\left\langle {
			{\textbf{u}''}^2
		} \right\rangle}
		\left\langle {
			\left( {
				\tilde{\textbf{u}} \cdot \nabla
			} \right) \overline{{\textbf{u}}''{}^2}
		} \right\rangle.
	\label{eq:kappa_nonequib_exp}
\end{eqnarray}
Here, the triple correlation terms of dispersion fluctuations have been dropped. On the right-hand side of the second equality in (\ref{eq:kappa_nonequib_exp}), the first and third terms represent how much the time-averaged coherent and incoherent fluctuation energies change in time along the coherent or plume motion $\tilde{\bf{u}}$, respectively. These terms are important since they directly reflect the non-equilibrium effect due to plume motions. On the other hand, the second and fourth terms contain correlations of the Lagrange derivatives along the coherent flow velocity $\tilde{\bf{u}}$ with the reciprocals of coherent and incoherent dissipation, respectively. These terms are expected to be small as compared to the former terms. Hence we dropped them in the second final line of (\ref{eq:kappa_nonequib_exp}).

	In the time--space double averaging procedure, the space average of the time-averaged turbulent energies are defined as
\begin{equation}
	\tilde{K} = \langle {\tilde{k}} \rangle
	= \left\langle {\overline{\tilde{\textbf{u}}^2}} \right\rangle/2
	= \left\langle {\tilde{\textbf{u}}^2} \right\rangle/2,
	\label{eq:disp_K_def}
\end{equation}
\begin{equation}
	K'' = \langle {k''} \rangle
	= \left\langle {\overline{{\textbf{u}}''{}^2}} \right\rangle/2
	= \left\langle {{\textbf{u}}''{}^2} \right\rangle/2,
	\label{eq:rand_K_def}
\end{equation}
\begin{equation}
	K = \langle {k} \rangle
	= \left\langle {\overline{{\textbf{u}}'{}^2}} \right\rangle/2
	= \left\langle {{\textbf{u}}'{}^2} \right\rangle/2.
	\label{eq:total_K_def}
\end{equation}

In (\ref{eq:kappa_nonequib_exp}), $\tilde{\tau} / \langle {\tilde{\textbf{u}}^2} \rangle$ and $\tau'' / \langle {{\textbf{u}}''{}^2} \rangle$ correspond to the reciprocals of dissipation rates of the coherent and incoherent fluctuation energies, respectively as
\begin{equation}
	\tilde{\varepsilon} 
	= {\left\langle {\tilde{\textbf{u}}^2} \right\rangle}/{(2\tilde{\tau})}
	\;\;\; \mbox{and}\;\;\;
	\varepsilon'' 
	= {\left\langle {{\textbf{u}}''{}^2} \right\rangle}/{(2\tau'')}.
	\label{eq:disp_eps_rand_eps}
\end{equation}

	We assume that the intensities of coherent and incoherent fluctuations, $\langle {\tilde{\bf{u}}^2} \rangle$ and $\langle {{\bf{u}}''{}^2} \rangle$, are comparable to each other with (\ref{eq:time_ave_en_rel}) as
\begin{equation}
	\langle{\tilde{\bf{u}}^2}\rangle
	\simeq \langle{{\bf{u}}''{}^2}\rangle
	\simeq \langle {{\bf{u}}'{}^2} \rangle/2.
	\label{similar_K}
\end{equation}
There are several situations where this assumption is reasonably well. For instance, in the situation considered in the following section, the surface cooling drives the descending motion of plumes. As the plume descends and finally disappears, the energy of the plume motions (coherent fluctuations) is transferred to the energy of random noises (incoherent fluctuations). This basic physics suggests the energy of the coherent and incoherent fluctuation energies are comparable. Actually, our direct numerical simulations show these two energies are comparable.

	In this case, we approximate (\ref{eq:kappa_nonequib_exp}) as
\begin{Eqnarray}
	\kappa_{\rm{N}} 
	&=& \tilde{C} \frac{1}{2 \tilde{\varepsilon}}
	\left\langle {
		(\tilde{\bf{u}} \cdot \nabla) \overline{\tilde{\bf{u}}^2}
	} \right\rangle
	+ C'' \frac{1}{2 \varepsilon''}
	\left\langle {
		(\tilde{\bf{u}} \cdot \nabla) \overline{{\bf{u}}''{}^2}
	} \right\rangle
	\nonumber\\
	&\simeq& C' \left( {
    	\frac{1}{\tilde{\varepsilon}} + \frac{1}{\varepsilon''}
	} \right) 
	\left\langle {
		(\tilde{\bf{u}} \cdot \nabla) \overline{{\bf{u}}'{}^2}
	} \right\rangle,
	\label{eq:kappa_noneq_approx}
\end{Eqnarray}
where the numerical factors are absorbed into the constant $C'$.

	As the averaging time relation (\ref{eq:ave_time_cond}) suggests, the characteristic time for the coherent fluctuation, $\tilde{\tau}$, is expected to be much longer than the counterpart for the incoherent fluctuation, $\tau''$. Under the condition of (\ref{similar_K}), the dissipation rates of the coherent and incoherent fluctuations may satisfy
\begin{equation}
	\tilde{\varepsilon} 
	= \frac{\langle {\tilde{\bf{u}}^2} \rangle}{\tilde{\tau}}
	\ll \frac{\langle {{\bf{u}}''{}^2} \rangle}{\tau''}
	= \varepsilon''.
	\label{eq:much_small_eps_coh}
\end{equation}
	Then, (\ref{eq:kappa_noneq_approx}) is reduced to
\begin{equation}
	\kappa_{\rm{N}}
	= \hat{C} \frac{1}{\tilde{\varepsilon}} \left\langle {
		(\tilde{\bf{u}} \cdot \nabla) \overline{{\bf{u}}'{}^2}
	} \right\rangle,
	\label{eq:kappa_noneq_Lambda}
\end{equation}
which implies that the non-equilibrium correction is expressed by the kinetic energy variation along the plume motion divided by the coherent energy dissipation rate.

	With the eddy-diffusivity expression (\ref{eq:kappa_noneq_Lambda}), the turbulent mass flux can be modelled as
\begin{Eqnarray}
	\langle {\overline{\rho' {\textbf{u}}'}} \rangle
	&=& - \kappa_{\textrm{E}} \left( {
		1 
		- \hat{C} \frac{1}{\tilde{\varepsilon}} 
		\left\langle {\frac{\tilde{D}k}{\tilde{D}t}} \right\rangle
	} \right) \nabla \left\langle {\overline{\rho}} \right\rangle
	\nonumber\\
	&\simeq& - \kappa_{\textrm{E}} \left( {
		1 
		- \hat{C} \frac{1}{\tilde{\varepsilon}} 
		\left\langle {(\tilde{\bf{u}} \cdot \nabla)k} \right\rangle
	} \right) \nabla \left\langle {\overline{\rho}} \right\rangle
	\label{eq:turb_mass_flux_kappa_noneq}
\end{Eqnarray}
where $\tilde{\varepsilon} (= \langle {\tilde{\bf{u}}^2} \rangle / \tilde{\tau})$ is the dissipation rate of the coherent fluctuation energy, $\tilde{D}/\tilde{D}t (= \partial/\partial t + \tilde{\bf{u}} \cdot \nabla)$ is the Lagrange derivative along the dispersion velocity $\tilde{\bf{u}}$, and $\hat{C}$ is the model constant. Equation~(\ref{eq:turb_mass_flux_kappa_noneq}) implies that the eddy diffusivity may change from its equilibrium value due to the non-equilibrium effect. The eddy diffusivity is enhanced or suppressed depending on the variation of kinetic energy along the coherent motion of the flow.

\subsection{Structure of the time--space double averaging model}\label{sec:model_struc}
	In the turbulence modelling approach, turbulent fluxes (transports due to unresolved motions) are expressed in terms of the mean or resolved fields coupled with the turbulent transport coefficients. The turbulent transport coefficients should be expressed by a few quantities that properly represent the statistical properties of the turbulence. The simplest and most widely employed model for the transport coefficients is based on the mixing-length theory (MLT) approximation of the turbulence characteristic length \citep{sti2002}. A more elaborate modelling method is to adopt some appropriate turbulent statistical quantities linked to the length and time scales of the turbulence, and consider the evolution of the transport equations of the turbulent statistical quantities. The turbulent kinetic energy and its dissipation rate (or directly the eddy turn-over time) are representative turbulent statistical quantities.	
	
	 In the simplest gradient-diffusion approximation, the turbulent mass flux $\langle {\rho' {\bf{u}}'} \rangle$ is modelled as
\begin{equation}
	\langle {\rho' {\bf{u}}'} \rangle
	= - \kappa_{\rm{T}} \nabla \langle {\rho} \rangle,
	\label{eq:turb_mass_flux_kappaT}
\end{equation}
where the transport coefficient, the eddy diffusivity $\kappa_{\rm{T}}$, is expressed in a generic form in terms of the turbulent energy $K = \langle {{\bf{u}}'{}^2} \rangle/2$ and the eddy turn-over time $\tau$ as
\begin{equation}
	\kappa_{\rm{T}} 
	= {\cal{F}} \{ {\tau, K} \}
	= {\cal{F}} \{ {\tau, \langle {{\bf{u}}'{}^2} \rangle } \}.
	\label{eq:kappaT_gen}
\end{equation}
In the formulation with the non-equilibrium effect, the generic form includes the advection velocity $\langle {\bf{u}} \rangle$ through the Lagrangian derivative. Then (\ref{eq:kappaT_gen}) is modified to
\begin{equation}
	\kappa_{\rm{T}} 
	= {\cal{F}} \{ {\langle {{\bf{u}}} \rangle; \tau, K} \}
	= {\cal{F}} \{ {\langle {{\bf{u}}} \rangle; \tau, \langle {{\bf{u}}'{}^2} \rangle } \}.
	\label{eq:kappaT_ne_gen}
\end{equation}
	 
	  In the time--space double averaging procedure, the turbulent mass flux $\langle {\overline{\rho' {\textbf{u}}'}} \rangle$ has to be expressed in terms of the time-averaged quantities and their space averages with the time-averaged fluctuation velocity and its space average. Then, the generic form for the eddy diffusivity with the non-equilibrium effect, $\kappa_{\rm{NE}}$, may be written as
\begin{equation}
	\kappa_{\rm{NE}}
	= {\cal{F}} \left\{ {
		{\left\langle {\overline{{\bf{u}}}} \right\rangle}, 
		{\overline{{\bf{u}}}; \tau'', \tilde{\tau}, K, k}
	} \right\}
	= {\cal{F}} \left\{ {
		\langle{\overline{{\bf{u}}}\rangle, 
		\tilde{{\bf{u}}}; \tau'', \tilde{\tau}, 
		\left\langle {\overline{{\bf{u}}'{}^2}} \right\rangle},
		\overline{{\bf{u}}'{}^2}
	} \right\},
	\label{eq:kappa_generic_form}
\end{equation}
where $\langle {\overline{\bf{u}}} \rangle$, $\overline{\bf{u}}$ and $\tilde{\bf{u}}$ are included for the Lagrangian derivatives. In addition, the time scales of the incoherent and coherent fluctuations, $\tau''$ and $\tilde{\tau}$, are included in (\ref{eq:kappa_generic_form}). The turbulent flux is expressed by the gradient of the space averaged field $\langle {\overline{f}} \rangle$ coupled with the turbulent transport coefficient. The transport coefficient is expressed in terms of the space fluctuation fields, i.e., the random/incoherent fluctuation $f''$ and the dispersive/coherent fluctuation $\tilde{f}$. In order to express the turbulent transport coefficient, we need information on the coherent fluctuation.

	The time-averaged total turbulent energy consists of the time-averaged coherent and incoherent fluctuation energies as (\ref{eq:time_ave_en_rel}), and the total turbulent energy (\ref{eq:total_K_def}) consists of the coherent fluctuation energy (\ref{eq:disp_K_def}) and the incoherent fluctuation energy (\ref{eq:rand_K_def}) as
\begin{subequations}\label{eq:space_ave_en}
\begin{equation}
	K
	= \tilde{K}
	+ K'',
	\label{eq:space_ave_K}
\end{equation}
or equivalently  
\begin{equation}
	\left\langle {\overline{{\bf{u}}'{}^2}} \right\rangle
	= \left\langle {\overline{{\tilde{{\bf{u}}}^2}}} \right\rangle
	+ \left\langle {\overline{{\bf{u}}''{}^2}} \right\rangle.
	\label{eq:space_ave_u_squared}
\end{equation}
\end{subequations}
Noting (\ref{eq:time_ave_en_rel}) and (\ref{eq:space_ave_en}), we see from (\ref{eq:kappa_noneq_Lambda}) that the non-equilibrium eddy diffusivity $\kappa_{\rm{NE}}$ is written as
\begin{equation}
	\kappa_{\textrm{NE}}
	= \left\{ { \everymath{\displaystyle}
	\begin{array}{lll}
	\kappa_{\textrm{E}} \left[ {
		1
		- C_{\tilde{\tau}} \frac{\tilde{\tau}}{\langle{{\textbf{u}}'{}^2}\rangle}
			\tilde{\Gamma}_D
	} \right]
	& \mbox{for}
	& \tilde{\Gamma}_D < 0,\\
	\rule{0.ex}{5.ex}
	\kappa_{\textrm{E}} \left[ {
		1
		+ C_{\tilde{\tau}} \frac{\tilde{\tau}}{\langle{{\bf{u}}'{}^2}\rangle}
			\tilde{\Gamma}_D
	} \right]^{-1}
	& \mbox{for}
	& \tilde{\Gamma}_D > 0,
	\end{array}
	} \right.
	\label{eq:kappa_noneq_form}
\end{equation}
with
\begin{equation}
	\tilde{\Gamma}_D 
	= \left\langle {
		(\tilde{\bf{u}} \cdot \nabla) \overline{{\bf{u}}'{}^2}
	} \right\rangle,
	\label{eq:Gamma_D_def}
\end{equation}
where $\tilde{\tau}$ is the time scale of the coherent fluctuation, and $C_{\tilde{\tau}}$ is the model constant. Again, we have approximate $\langle {\overline{\bf{u}}} \rangle \simeq \tilde{\bf{u}}$ in the Lagrangian derivative since $\langle {\overline{\bf{u}}} \rangle \simeq 0$. In (\ref{eq:kappa_noneq_form}) we adopt a simple Pad\'{e} approximation in order to avoid unphysical negative diffusivity.

	Equation~(\ref{eq:kappa_noneq_form}) implies that the non-equilibrium effect arising from the variation along the coherent velocity motion $\tilde{\bf{u}}$ will alter the effective turbulent diffusivity $\kappa_{\rm{NE}}$ as compared with the equilibrium counterpart $\kappa_{\rm{E}}$.

\section{Application to stellar convection}\label{sec:stell_conv}
As was referred to in \S~\ref{sec:intro}, it has been argued that the surface cooling and the resulting down-flowing plumes play an important role in turbulent mixing, by altering the turbulence statistics in the stellar convection zone \citep{spr1997,cos2016}. A turbulence model with the simplest mixing-length theory (MLT) should be modified for such convective flows \citep{bra2016}. In this section, we apply the eddy-diffusivity expression with the non-equilibrium effect in the time--space double averaging procedure, (\ref{eq:kappa_noneq_form}), to a stellar convection problem.

\begin{figure}
\centering
\includegraphics[width= \columnwidth]{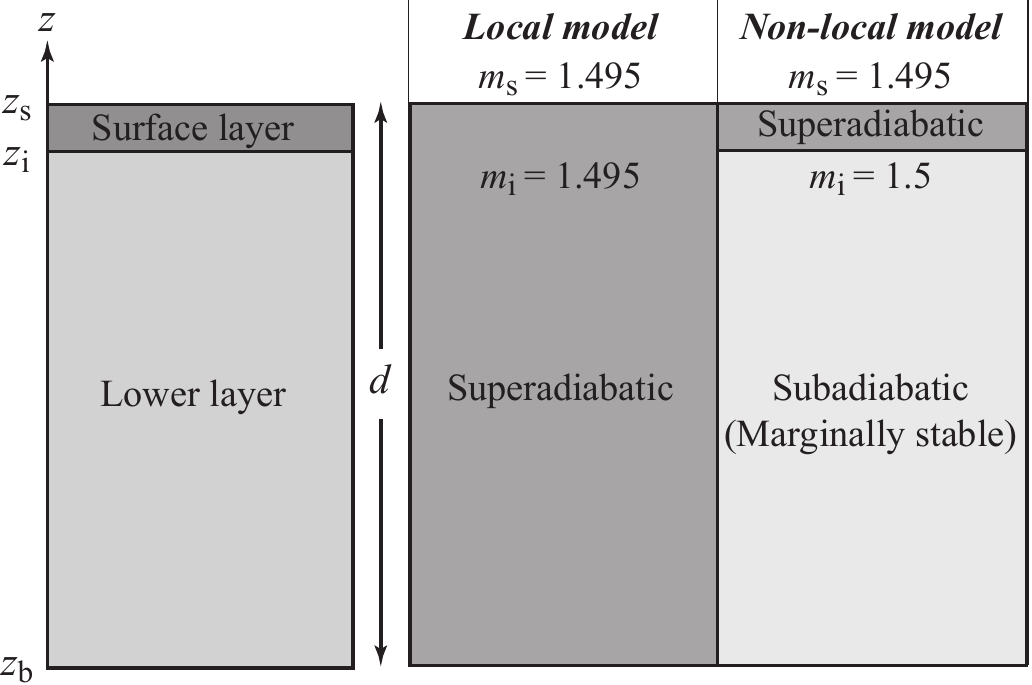}
\caption{Two-layer polytropic gas configuration. The surface layer ($z_{\rm{i}} \le z \le z_{\rm{s}}$) and the lower layer ($z_{\rm{b}} \le z \le z_{\rm{i}}$) of the convection zone. In the local model convection is driven by the weakly superadiabatic entropy gradient across the full depth $d$ of the convection zone ($z_{\rm{b}} \le z \le z_{\rm{s}}$) with the polytropic index $m_{\rm{s}} = m_{\rm{i}} = 1.495$ (superadiabatic), while in the non-local model convection is driven by cooling at the surface layer ($z_{\rm{i}} \le z \le z_{\rm{s}}$) with $m_{\rm{s}} = 1.495$ (superadiabatic) and $m_{\rm{i}} = 1.5$ (marginally stable).}
          \label{fig:two-layer}
        \end{figure}

	We simulate stellar convection by solving the compressible hydrodynamic equations (\ref{eq:den_eq})-(\ref{eq:int_en_eq}). Following the set-up mimicking a local domain of the stellar convection zone in \citet{cos2016}, we consider a computational domain in a rectangular box, where $z$ is the vertical coordinate, which directs upward from the bottom $z_{\rm{b}}$ to the top surface $z_{\rm{s}}$ ($z_{\rm{b}} \le z \le z_{\rm{s}}$). The acceleration of gravity is downward or negative $z$ direction. The horizontal coordinates are $x$ and $y$, and the horizontal boundaries are periodic. We adopt a set-up based on the two-layer polytropic gas convection, where the upper portion of the domain ($z_{\rm{i}} \le z \le z_{\rm{s}}$) is the surface layer and the lower portion ($z_{\rm{b}} \le z \le z_{\rm{i}}$) is the residual layer (Fig.~\ref{fig:two-layer}). In the present calculation, the upper $5 \%$ portion is set as the surface layer [$z_{\rm{i}}/d = 0.95$, $d (= z_{\rm{s}} - z_{\rm{b}})$: full depth of the convection zone].

	We assume a polytropic relation between the pressure and density as $p = \rho^{1+1/m}$ with the polytropic index $m$. In the adiabatic case, $p = \rho^\gamma$ with $\gamma (= C_p/C_v)$ being the ratio of the specific heat at constant pressure $C_p$ to that at constant volume $C_v$. The hydrostatic balance is also assumed for the equilibrium state. With these assumptions, the spatial distributions of density $\rho$ and pressure $p$ are determined by the specific internal energy $e$ at the top surface. For the two-layer polytropic gas model, we consider two cases. The first one is {\it the local model}, where convection is driven by a weakly superadiabatic ($m_{\rm{s}} = m_{\rm{i}} = 1.495 < 1.5$, $m_{\rm{s}}$: polytropic index at the top surface, $m_{\rm{i}}$: polytropic index at the top of the residual layer) local entropy gradient throughout the convection zone. The other one is {\it the non-local model}, where convection is driven by a surface cooling, only the vicinity of the surface ($0.95 \le z/d \le 1$) is superadiabatic. The surface layer is convectively unstable with $m_{\rm{s}} = 1.495 < 1.5$, and the lower residual layer ($0 \le z/d \le 0.95$) is marginally stable with $m_{\rm{i}} = 1.5$. 
	
	The non-dimensional parameters in our simulation; the Prandtl number is $Pr \simeq 1$ and the Rayleigh number is $Ra \simeq 4.2 \times 10^{6}$. The density contrast between the bottom and top surface is $\rho_{\rm{b}} /\rho_{\rm{s}} \simeq 100$ in both cases. In order to sustain the superadiabaticity at the surface in the cooling-driven case, a Newtonian-cooling term is added in the energy equation, following \citet{cos2016}.
Details of the numerical simulations including the set-up (the initial density and pressure distributions, the values of physical parameters, etc.) as well as the arguments on other turbulent fluxes, such as the turbulent internal-energy flux $\langle {e' {\bf{u}}'} \rangle$, are given in our Paper II.
	
\begin{figure}
\includegraphics[width=\columnwidth]{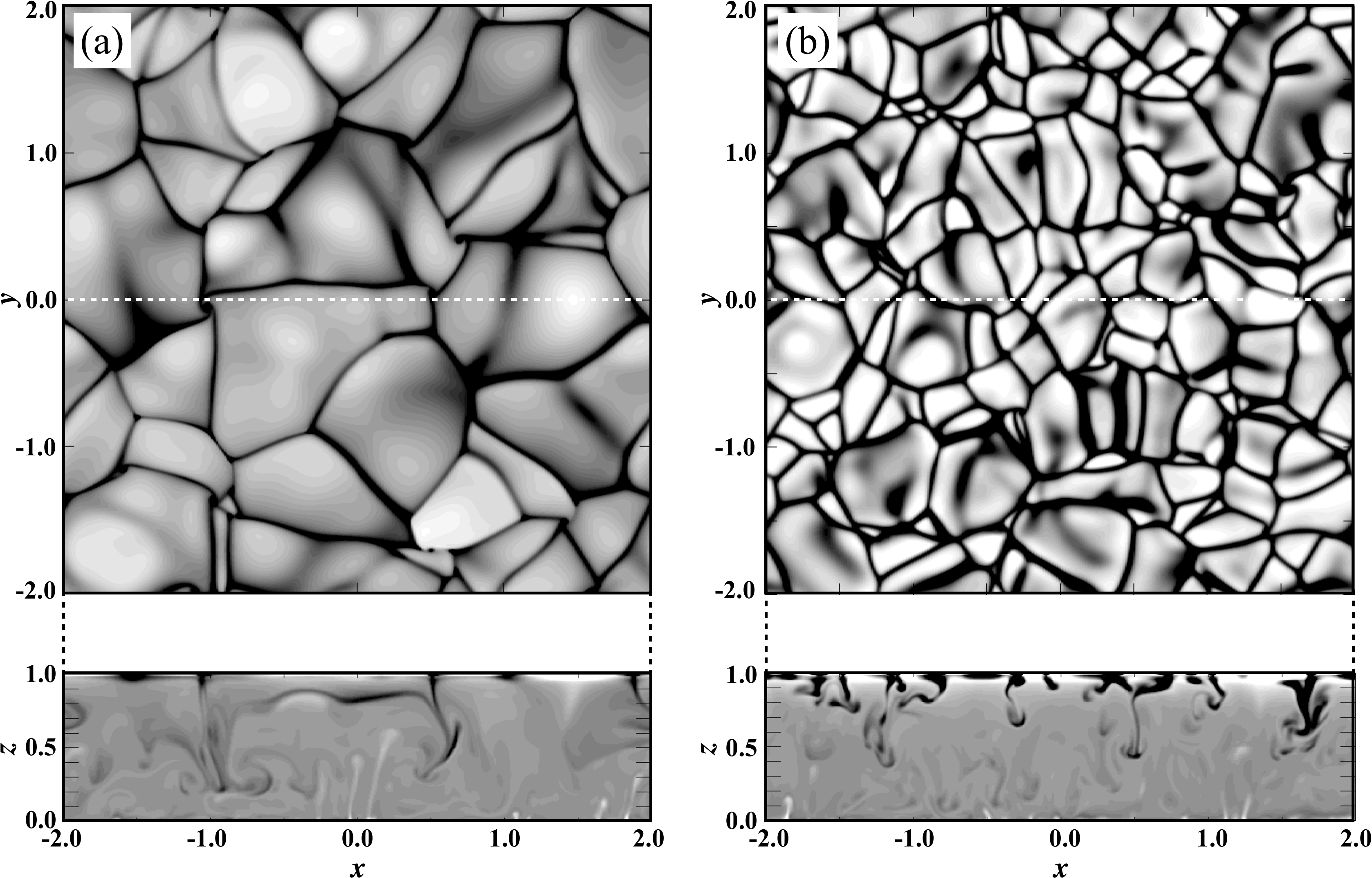}
\caption{Entropy distributions in our direct numerical simulations (DNSs) for the locally-driven case (a) and the non-locally driven case (b). The horizontal cross-sections of the entropy fluctuation $s' (= s - \langle{s}\rangle)$ at the top surface (Top). In the non-locally- or cooling-driven case, the horizontal extension of the cell structures is much more limited than the counterpart in the locally-driven case. The vertical cross-sections of the entropy fluctuation $s'$ from the horizontal mean (Bottom). In the non-locally driven case, the low entropy down-flow or plume structures produced at the surface are prominent in the upper region.}
    \label{fig:entropy_contour_dns}
\end{figure}

	To see the qualitative differences between the local and non-local models, we show, in Figure~\ref{fig:entropy_contour_dns}, the convection profile for each model at a quasi-steady state. The distributions of the entropy fluctuation, $s' (= s - \langle {s} \rangle)$, at the top surface layer (top panel) and the vertical cutting plane (bottom panel) are demonstrated for the (a) local model and (b) non-local model, where the angular brackets denotes the horizontal average. The black tone corresponds to the downflow region with a negative entropy fluctuation (i.e., negative buoyancy). There exist a lot of downward plumes in the upper part of the convection zone in the
cooling-driven model, while the local model shows less plumes and the development of a larger convective structure.

\subsection{Entropy-gradient driven local transport and cooling driven non-local transport}
\begin{figure}
  \includegraphics[width=\columnwidth]{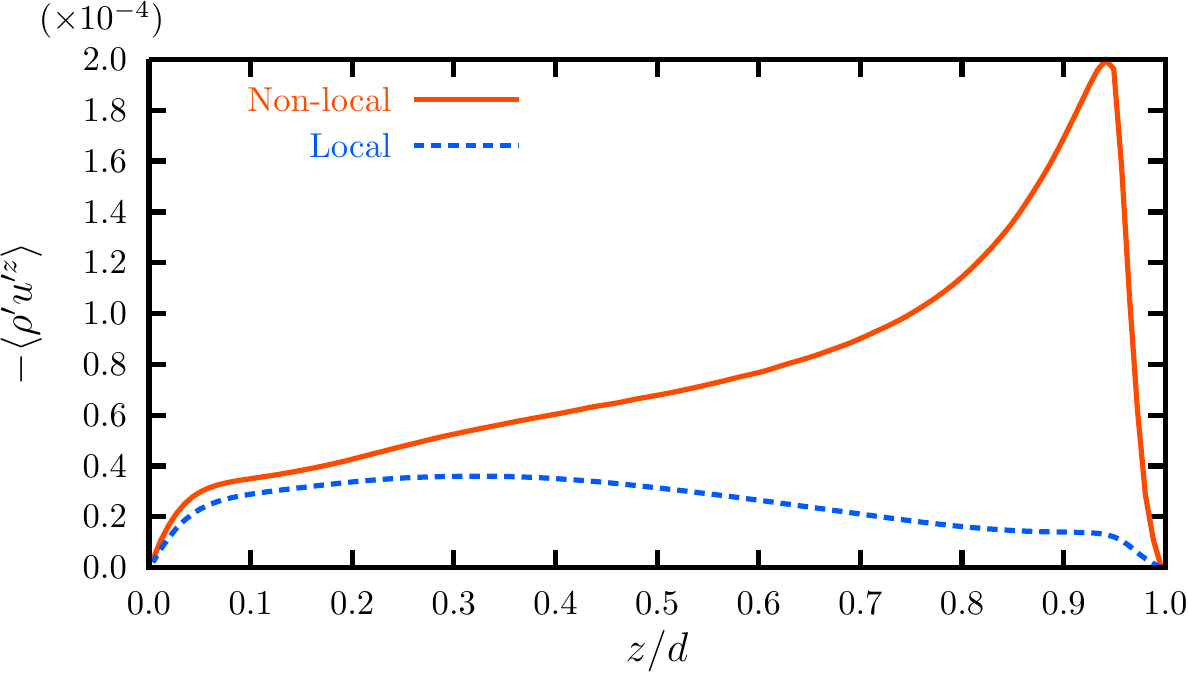}
  \caption{Spatial profile of the turbulent mass flux $\langle {\rho' u'{}^z} \rangle$ in our direct numerical simulations (DNSs) for the locally- and non-locally driven cases. In locally-driven case, the turbulent mass flux monotonically increases with the depth except in the vicinity of the bottom of the convection zone. In the non-locally- or cooling-driven case, the turbulent mass flux has a strong peak in the near surface region, and monotonically decreases with the depth.}
  \label{fig:turb_mass_flux_dns}
\end{figure}

According to direct numerical simulations (DNSs), the spatial profile and amplitude of the turbulent mass flux $\langle {\rho' {\bf{u}}'} \rangle$ are fairly different between the entropy-gradiend-driven local mixing and the cooling-driven non-local mixing (Fig.~\ref{fig:turb_mass_flux_dns}). If the convective motion is driven by the weakly superadiabatic ambient state across the full depth of the convection zone (locally-driven case), the turbulent mass flux increases monotonically with the depth except in the vicinity of the bottom of the convective zone (the dashed or ``local'' plot in Fig.~\ref{fig:turb_mass_flux_dns}). On the other hand, if the convective motion is driven by the superadiabatic state confined to the upper layer in the vicinity of the surface (cooling- or non-locally-driven case), the turbulent mass flux shows a strong peak at a shallow region near the surface ($z \sim 0.9 d$), then monotonically decreases with the depth except in the vicinity of the bottom (the solid or ``non-local'' plot in Fig.~\ref{fig:turb_mass_flux_dns}). The amplitude of the turbulent mass flux at $z \sim 0.9 d$ in the cooling-driven case is about one order higher than that in the locally-driven case.

\begin{figure}
\centering
\includegraphics[width= \columnwidth]{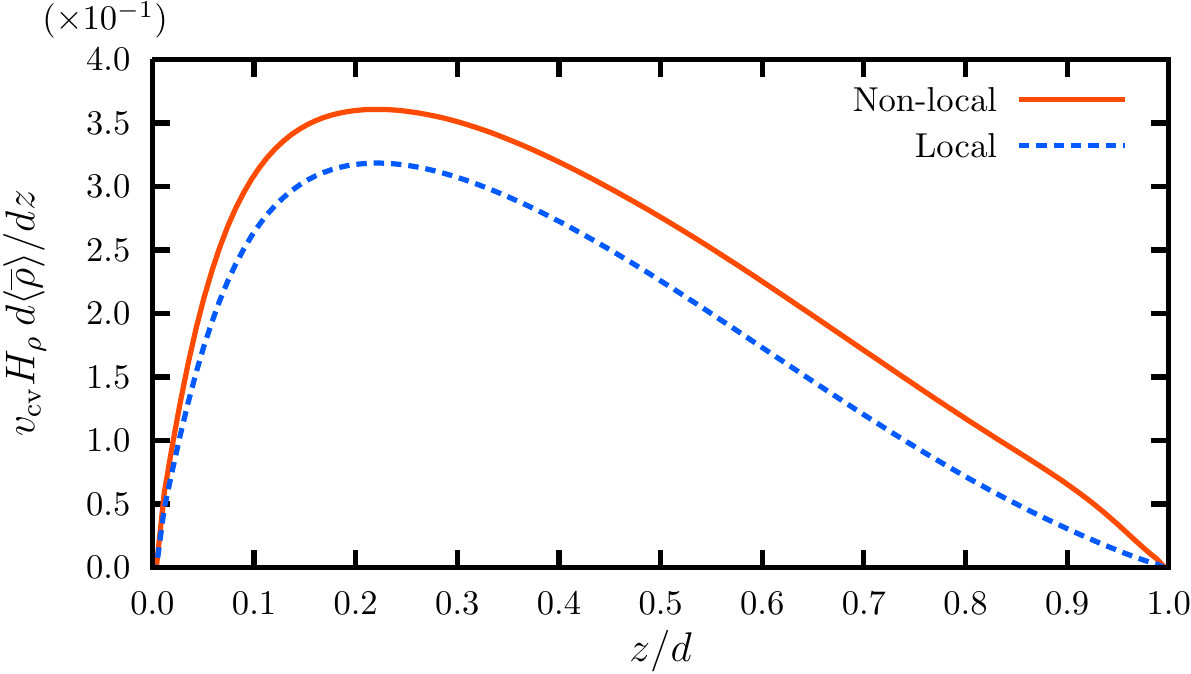}
\caption{Spatial profile of the gradient diffusion model with MLT. The spatial distribution of the mean density gradient $d\langle {\overline{\rho}} \rangle/dz$ multiplied by the characteristic convective velocity $v_{\rm{cv}}$ and the density scale height $H_\rho = |dz/d\ln \langle {\overline{\rho}} \rangle|$ is plotted.}
\label{fig:grad_transport_model_mlt}
\end{figure}

	The gradient diffusion model with MLT is expressed as
\begin{equation}
	\langle {\rho' u'{}^z} \rangle_{\rm{MLT}}
	\sim - v_{\rm{cv}} H_\rho 
		\frac{d \langle {\overline{\rho}} \rangle}{dz},
	\label{eq:mlt_model_turb_mass_flux}
\end{equation}	
where $v_{\rm{cv}} = \left\langle {(\overline{u'{}^z)^2}} \right\rangle^{1/2}$ is the characteristic convective velocity, and $H_\rho$ is the density scale height defined by 
\begin{equation}
	H_\rho 
	= \frac{\langle {\overline{\rho}} \rangle}
		{\left| {\partial \langle {\overline{\rho}} \rangle / 
			\partial z
		} \right|}
	= \left| {\frac{dz}{d\ln \langle {\overline{\rho}} \rangle}} \right|.
	\label{eq:den_scl_hght}
\end{equation}
The spatial distribution of (\ref{eq:mlt_model_turb_mass_flux}) is plotted in Fig.~\ref{fig:grad_transport_model_mlt}. Comparing Figs.~\ref{fig:turb_mass_flux_dns} and \ref{fig:grad_transport_model_mlt}, we see that the spatial profile of the gradient transport model with MLT (\ref{eq:mlt_model_turb_mass_flux}) (Fig.~\ref{fig:grad_transport_model_mlt}) is quite similar to that of $\langle {\rho' u'{}^z} \rangle$ for the locally-driven convection case in DNS (Fig.~\ref{fig:turb_mass_flux_dns}). Then, we see that the spatial profile of the turbulent mass flux $\langle {\rho' {\bf{u}}'} \rangle$ in the locally-driven case can be readily described by the gradient-diffusion-type model with a simple mixing-length theory (MLT) of the turbulent eddy diffusivity $\kappa_{\rm{T}}$.

	In contrast to the locally-driven case, the spatial profile and large amplitude of the turbulent mass flux in the cooling-driven or ``non-local'' case cannot be reproduced by such a simple gradient-diffusion type model. This is also the case for the turbulent internal-energy flux $\langle {e' {\bf{u}}'} \rangle$. The turbulent transport cannot be properly reproduced by a simple MLT model in the cooling-driven or ``non-local'' convection. In the following, we address this problem with the aid of a turbulence model with the plume effect incorporated through the non-equilibrium effect.

\subsection{Incoherent and coherent fluctuations\label{sec:incoh_coh_flucts}}
\begin{figure}
\includegraphics[width=\columnwidth]{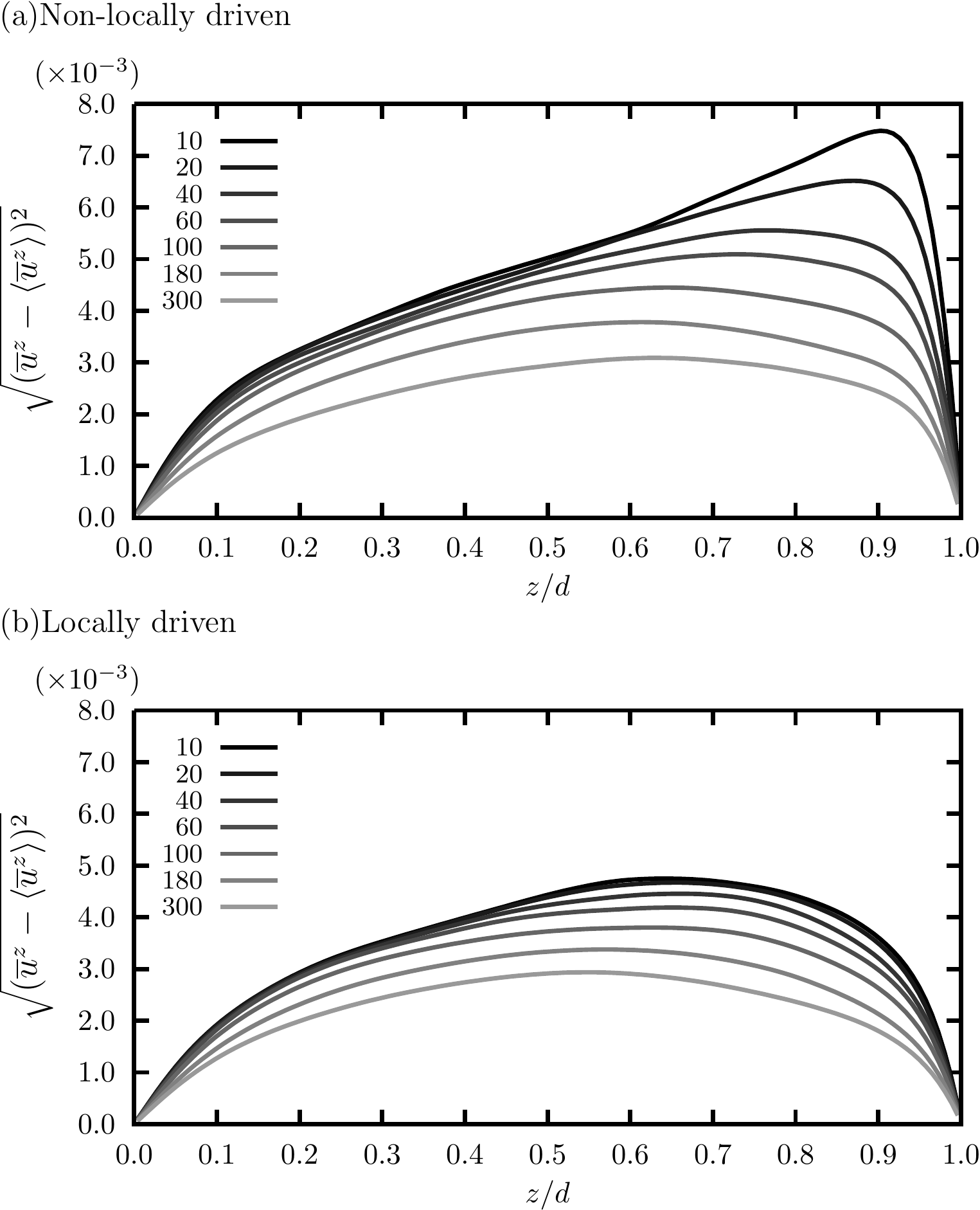}
\caption{Spatial profile of the amplitude of the coherent fluctuations $\sqrt{(\tilde{u}^z)^2} [= \sqrt{(\overline{u}^z - \langle {\overline{u}^z} \rangle)^2}]$ in the non-locally-driven case (a) and locally-driven case (b). The variations of the spatial profiles of $\sqrt{(\tilde{u}^z)^2}$ depending on the averaging time $T$ ($T = 10 - 300$) are also shown in both cases.}
    \label{fig:coh_incoh_fluct}
\end{figure}

In the time--space double averaging procedure, the coherent velocity fluctuation or dispersion $\tilde{\bf{u}}$ is given by
\begin{equation}
	\tilde{\bf{u}}
	= \overline{\bf{u}}
	- \langle {\overline{\bf{u}}} \rangle.
	\label{eq:coh_fluct_vel_def}
\end{equation}
As was referred to in \S~\ref{sec:double_ave}, the value of the coherent fluctuation sensitively depends on the window of time filtering or the averaging time $T$ in the definition of the time average (\ref{eq:time_ave_short}). If the averaging time $T$ is much longer than a typical plume lifetime, the coherent motions of plumes would be statistically cancelled out, and the magnitude of the coherent velocity would be considerably reduced. For a sufficiently long $T$, the magnitude of the coherent fluctuation would be negligible to the counterpart of the random or incoherent fluctuation. In Fig.~\ref{fig:coh_incoh_fluct}, the spatial distributions of the magnitude of the coherent velocity fluctuation $\sqrt{(\tilde{u}^z)^2} [= \sqrt{(\overline{u}^z - \langle {{\overline{u}}^z} \rangle)^2}]$ in the non-locally- and locally-driven cases with varying the averaging time $T$ are presented. In the cooling-driven case (Top), the amplitude of the coherent fluctuation with a short averaging time has an eminent peak in the near surface region ($z/d \sim 0.9$). This peak position is the same as that of the strong peak of the turbulent mass flux in the cooling-driven case as was shown in Fig.~\ref{fig:turb_mass_flux_dns}. This implies that the spatial distribution of the turbulent mass flux is determined by the coherent component of the fluctuation, which represents the plume motion. As $T$ is set longer, the amplitude of $|\tilde{\bf{u}}|$ becomes smaller. This decrease tendency is most prominent in the shallow region where a large peak  of the coherent fluctuation is located in the non-locally driven case. There as $T$ increases, the large values of $\sqrt{(\tilde{u}^z)^2}$ decreases, eventually the spatial profile peaked in the shallow region disappears and the spatial profile becomes more similar to the one in the locally-driven case. This clearly shows that the prominent characteristics of the non-locally- or cooling-driven convection motion, which are directly related to the plume motion, can be described by the coherent component of the fluctuating motion in the present time--space double averaging procedure. Note that from the scaled convective velocity $v^z = \sqrt{\langle{(u^z)^2}\rangle} \sim 0.01$ and the scaled full depth of the convective zone $d = 1$, the (maximum) lifetime of the plume might be estimated as $(d/2)/v^z \sim 50$, which corresponds to the number of snapshots of $25$. This suggests that we should set the averaging time $T \lesssim 25\ ({\rm{snapshots}})$. 

\subsection{Modelling of the enhanced transport due to the plume through the non-equilibrium effect}
Considering the relevance of the coherent fluctuation in the non-locally- or cooling-driven convection, we apply the non-equilibrium eddy-diffusivity model to the turbulent mass flux in a stellar convection zone.

\begin{figure}
  \centering
\includegraphics[width= 0.75  \columnwidth]{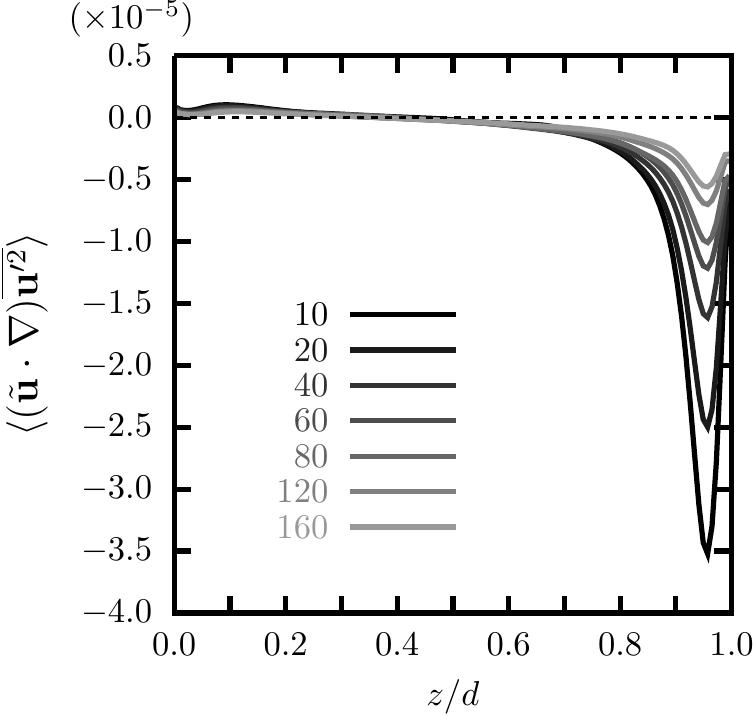}
\caption{Spatial profile of the non-equilibrium effect along the coherent fluctuation or dispersion velocity $\langle {(\tilde{\bf{u}} \cdot \nabla) \overline{{\bf{u}}'{}^2}} \rangle$ in (\ref{eq:kappa_noneq_form}) with various averaging time $T$ ($T = 10-160$).}
    \label{fig:noneq_effect}
\end{figure}

	With the aid of the DNSs, we evaluate the non-equilibrium effect associated with the coherent velocity, $\langle {(\tilde{\bf{u}} \cdot \nabla) {\overline{{\bf{u}}'{}^2}}} \rangle$, in (\ref{eq:kappa_noneq_form}). Figure~\ref{fig:noneq_effect} shows the spatial distribution of the non-equilibrium effect in the non-locally or cooling driven cases with various averaging time $T$. In accordance with the results of the magnitude of the coherent fluctuation presented in Fig.~\ref{fig:coh_incoh_fluct}, the non-equilibrium effect associated with the plume motion is prominent in the near surface region ($z/d \ge 0.9$) in the small $T$ cases ($T \lesssim 20$). Because of the cooling at the surface layer, the coherent velocity $\tilde{\bf{u}}$ is expected to be statistically dominant in the downward direction ($\tilde{u}^z < 0$). On the other hand, the turbulent energy $\overline{{\bf{u}}'{}^2}$ decreases in the downward direction as $\nabla \overline{u'{}^2} > 0$. Then, the non-equilibrium effect is expected to be negative as
\begin{equation}
	\langle {
		(\tilde{\bf{u}} \cdot \nabla) {\overline{{\bf{u}}'{}^2}}
	} \rangle
	= \left\langle {
		\left( {\tilde{u}^z \frac{\partial}{\partial z}} \right)
			{\overline{{\bf{u}}'{}^2}}
	} \right\rangle
	< 0.
	\label{eq:sign_ne_effect}
\end{equation}
It follows from (\ref{eq:kappa_noneq_form}) that the non-equilibrium eddy diffusivity $\kappa_{\rm{NE}}$ is enhanced as compared to the equilibrium eddy diffusivity $\kappa_{\rm{E}}$.

With the enhanced eddy diffusivity with the non-equilibrium effect included, $\kappa_{\rm{NE}}$ (\ref{eq:kappa_noneq_form}), the turbulent mass flux is evaluated. The turbulent mass flux $\langle {\overline{\rho' u'{}^z}} \rangle$ is given by
\begin{equation}
	\langle {\overline{\rho' u'{}^z}} \rangle
	= - \kappa_{\rm{NE}} 
		\frac{\partial \langle {\overline{\rho}} \rangle}{\partial z},
	\label{eq:turb_mass_flux_ne_model}
\end{equation}
where $\kappa_{\rm{NE}}$ is the eddy-diffusivity coefficient with the non-equilibrium effect included as in (\ref{eq:kappa_noneq_form}).

	In the present application, the non-equilibrium model is given as 
\begin{equation}
	\kappa_{\rm{NE}}
	= \kappa_{\rm{E}} \left[ {
		1 - C_{\tilde{\varepsilon}} 
			\frac{1}{\tilde{\varepsilon}} \left\langle {
				(\tilde{\bf{u}} \cdot \nabla) \overline{{\bf{u}}'{}^2}
			} \right\rangle
	} \right]
	\label{eq:nonequi_model_appl}
\end{equation}
with the equilibrium diffusivity $\kappa_{\rm{E}}$ being formulated by the mixing-length theory expression
\begin{equation}
	\kappa_{\rm{E}}
	= C_{\rm{m}} \left\langle {\overline{(u'{}^z)^2}} \right\rangle^{1/2}
		H_\rho
	\label{eq:kappa_e_mlt}
\end{equation}
[$H_\rho$: the density scale height defined by (\ref{eq:den_scl_hght})] and $C_{\tilde{\varepsilon}}$ and $C_{\rm{m}}$ model constants.

	In the expression of the non-equilibrium model (\ref{eq:nonequi_model_appl}), we need to evaluate the dissipation rate of the coherent fluctuation energy, $\tilde{\varepsilon}$. In this application, we express $\tilde{\varepsilon}$ in terms of the density of the ambient fluid as
\begin{equation}
	\tilde{\varepsilon} = \langle {\overline{\rho}} \rangle^p
	\label{eq:coh_eps_rho_p}
\end{equation}
with $p$ being an appropriate index, whose value shall be evaluated with the aid of the plume dynamics.

	In Appendix~\ref{sec:append_C}, it is shown that the vertical flux of the plume energy can be expressed in terms of the buoyancy flux as in (\ref{eq:en_flux_buoy_rel}). This suggests that the dissipation rates of the coherent fluctuation, $\tilde{\varepsilon}$, can be estimated as
\begin{equation}
	\tilde{\varepsilon} 
	\sim \frac{(\tilde{u}^z)^2}{\tilde{\tau}}
	\sim \frac{(\tilde{u}^z)^3}{\ell_z}
	\sim \frac{B}{b^2},
	\label{eq:tilde_eps_B_rel}
\end{equation}
where $\ell_z$ is the vertical length scale associated with the plume motion, $B$ is the buoyancy flux (defined by (\ref{eq:buoy_flux_def})), and $b$ is the lateral extension length scale of the plume.	We see from (\ref{eq:tophat_b_exp}) that the lateral extension of the plume, $b$, is basically related to the volume flux $Q$ (defined by (\ref{eq:volume_flux_def})) as
\begin{equation}
	b^2 \propto Q^2.
	\label{eq:plume_b_Q_rel}
\end{equation}
Using (\ref{eq:Q_B_scaling}) for the relationship between $Q$ and $B$, we have
\begin{equation}
	\tilde{\varepsilon}
	\sim \frac{B}{Q^2}
	\sim B^{1/3}.
	\label{eq:plumetilde_eps_B_rel}
\end{equation}
as one of the simplest approximations.

	It follows from (\ref{eq:buoy_flux_def}) [or (\ref{eq:weight_deficiency_def})] that the dependency of the buoyancy flux $B$ on the ambient or environmental density $\rho_e$ is expressed as
\begin{equation}
	B \sim \rho_e.
	\label{eq:plume_B_rho_e_rel}
\end{equation}
With (\ref{eq:plumetilde_eps_B_rel}), this suggests that we can put $p = 1/3$ in (\ref{eq:coh_eps_rho_p}). Thus, we finally get a simple model expression of the non-equilibrium turbulent diffusivity as
\begin{equation}
	\kappa_{\rm{NE}}
	= \kappa_{\rm{E}} \left[ {
		1 - C \langle{\overline{\rho}}\rangle^{-1/3} \left\langle {
				(\tilde{\bf{u}} \cdot \nabla) \overline{{\bf{u}}'{}^2}
			} \right\rangle
	} \right]
	\label{eq:simple_noneq_model}
\end{equation}
with the model constant $C$. Of course, this result is on the basis of crude approximations. So, the dependence of $\tilde{\varepsilon}$ on $\langle {\rho} \rangle$ in (\ref{eq:coh_eps_rho_p}) might be with a different positive $p$ other than $p=1/3$.

\begin{figure}
\centering
\includegraphics[width=\columnwidth]{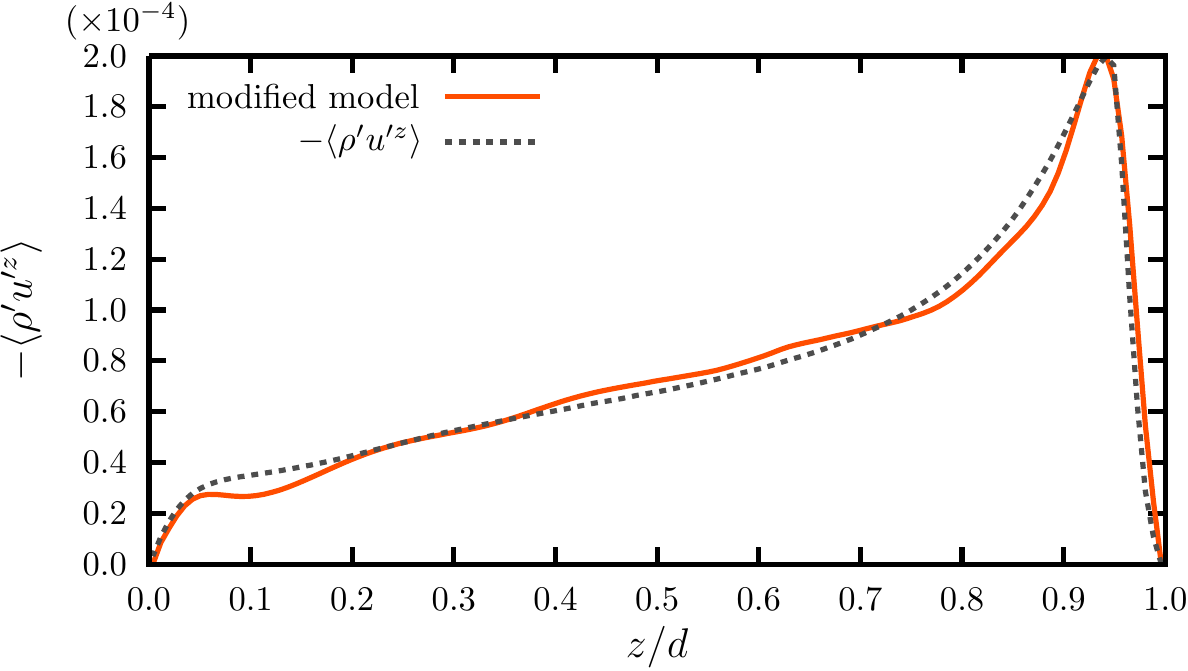}
          \caption{Spatial profiles of the turbulent mass flux $\langle {\overline{\rho' {\bf{u}}'}} \rangle$ in the non-locally-driven case. The solid or ``non-local'' line represents the DNS result. The dashed or ``model'' line represents the present model (\ref{eq:turb_mass_flux_ne_model}) with the averaging time $T=20$. The equilibrium eddy diffusivity is expressed by the mixing length model $\kappa_{\rm{E}} = \left\langle {\overline{(u'{}^z)^2}} \right\rangle^{1/2} H_\rho$, (\ref{eq:kappa_e_mlt}), with the density scale height $H_\rho = \langle {\overline{\rho}} \rangle / (\partial \langle {\overline{\rho}} \rangle / \partial z) = dz/d \ln \langle {\overline{\rho}} \rangle$, (\ref{eq:den_scl_hght}).}
          \label{fig:noneq_turb_mass_flux}
        \end{figure}

	The spatial profile of the turbulent mass flux $\langle {\overline{\rho' u'{}^z}} \rangle$ is presented in Fig.~\ref{fig:noneq_turb_mass_flux}. Here, $\langle {\overline{\rho' u'{}^z}} \rangle$ expressed by the non-equilibrium model (\ref{eq:turb_mass_flux_ne_model}) with $\kappa_{\rm{NE}}$ given by (\ref{eq:simple_noneq_model}) and the averaging time $T=20$ (dashed) should be compared with $\langle {\overline{\rho' u'{}^z}} \rangle$ obtained from the DNS (solid).

	We see from Fig.~\ref{fig:noneq_turb_mass_flux} that the result of the present non-equilibrium model basically agrees with the result of DNS. In particular, we can reproduce a peak in the shallow region ($z/d \sim 0.9$) and the general decreasing tendency of the magnitude with the depth $z$, which cannot be reproduced at all by using the eddy diffusivity $\kappa_{\rm{E}}$ with a simple mixing-length model (\ref{eq:kappa_e_mlt}).

	As was referred to in the final paragraph in \S~\ref{sec:m-field_turb_correl}, in the self-consistent turbulence model, the expressions of the turbulent coefficients should be determined by the nonlinear dynamics of the mean and fluctuation fields without resorting to the externally determined transport coefficients. However, in this work, for the sake of simplicity, the simplest model expression based on the MLT approximation for the equilibrium eddy diffusivity $\kappa_{\rm{E}}$ (\ref{eq:kappa_e_mlt}) is employed. In this sense, the present model is not fully self-consistent. It is possible to use the present non-equilibrium formulation in a more elaborated framework such as the $K-\ell$ model or  the $K-\varepsilon$ model, where the equation of the length-scale $\ell$ or the dissipation-rate $\varepsilon$ as well as the equation of the turbulent energy will be employed to construct a closed system of equations. Using such an elaborated self-consistent framework of the turbulence model, without resorting to the MLT hypothesis, would be an interesting subject for the next step.

\section{Concluding remarks}\label{sec:concl}

The effects of plume on the convective turbulent flux are incorporated into a mean-field model through the non-equilibrium effect on the transport coefficients. In order to extract the variation of turbulence along a plume motion, a time--space double averaging procedure is adopted. With this double averaging, the fluctuation is divided into the coherent and incoherent components. The turbulent energy conversions between the coherent and incoherent components are examined with the aid of the evolution equations of both fluctuation components. In this framework, the turbulent transport coefficients are expressed in terms of time- and space-averaged turbulent energies. 

	Without introducing the time averaging procedure, the space or ensemble average of the velocity is just small as $\langle {u^z} \rangle \simeq 0$, so that the non-equilibrium effect  cannot be implemented into a mean-field model in the simplest space or ensemble averaging framework. Using the mean velocity associated with a higher-order correlation such as the variance $\langle {u^z{}^2} \rangle (\equiv \sigma^2)$, the skewness $\langle {u^z{}^3} \rangle / \sigma^3$, the kurtosis $\langle {u^z{}^4} \rangle / \sigma^4$, etc.\ may be possible. In particular, one may consider that the skewness is relevant for extracting the effects of plume since the skewness reflects the lack of symmetry between down- and up-flows associated with the descending and ascending plume motions. However, according to our DNSs (the results are not shown in this paper), as long as the space or ensemble averaging is adopted, the non-equilibrium effect due to the variation along the skewness-related velocity is not localised in the shallow region, where the plume effect is predominant, but is broadly distributed in the full depth of the convection zone. In this sense, the skewness is not so much relevant to the non-equilibrium effect. Rather, we are required to use the coherent velocity introduced by the time--space averaging. 

	The essential ingredients of the present non-equilibrium effect are the coherent fluctuating motion representing the plume, and the inhomogeneities of turbulence (turbulent energy and its dissipation rate) along the coherent flow. By the non-equilibrium effect, the time and length scales of turbulence are altered as compared with the counterparts in the equilibrium case, and the turbulent transport is enhanced or suppressed depending on the properties of the variation along the coherent motion.

	This non-equilibrium model was applied to a stellar convection driven by cooling at the surface layer to evaluate the turbulent fluxes. The turbulent mass flux in the cooling-driven convection is characterised by its enhanced and localised spatial distribution just below the cooling surface. This property is marked contrast with the counterpart in the convection locally driven by weakly superadiabatic ambient state across the full depth of the convective zone. This profile of the cooling- or non-locally-driven convection cannot be reproduced at all by the usual eddy-diffusivity model with MLT expression of the turbulent transport coefficient.
	
	With setting the averaging time $T$ for the time average similar or less than a typical estimated lifetime of plumes, we observe a large amount of the coherent fluctuations, especially in the shallow domain. This indicates that plume motions can be detected as coherent fluctuations in the time--space double averaging procedure.
	
	The location of strong coherent motions coincides with the peak position of the turbulent mass flux in the direct numerical simulations (DNSs). This indicates that the non-equilibrium effect associated with the variation along the plume motion is strongly related to the enhancement of turbulent transport in the non-locally- or cooling-driven convection.

	With the non-equilibrium turbulence effect, the spatial distribution of the turbulent mass flux was successfully reproduced by the present transport model. This non-equilibrium model can also reproduce the spatial distribution of the turbulent internal-energy flux in the cooling-driven stellar turbulent convection. This point as well as the details of numerical set-up (values of physical parameters, initial distribution of pressure, density, etc.) and numerical results will be reported in our Paper II.
	
	In the present simulation, the turbulent mass flux predicted from the non-equilibrium model is compared with the true turbulent mass flux that is obtained from direct numerical simulation (DNS). This is the so-called {\it a priori} test of turbulence model, where the true values of the velocity and turbulent energy calculated by DNS are used. We see from test that the present turbulence model with the effects of plume incorporated through the non-equilibrium expression on the turbulent transport works well in describing the turbulent transport in the cooling-driven convection. It is expected that the non-equilibrium turbulence model in the time--space double averaging procedure can extend the scope of the simple mean-field turbulence modelling approach in the stellar convective turbulence.
	
	As was referred to at the end of \S~\ref{sec:stell_conv}, constructing a turbulence model by utilising the transport equations of the turbulent statistical quantities such as the turbulent energy and its dissipation rate, $K$ and $\varepsilon$, without resorting to the externally determined parameters ($v^z$, $H_\rho$), is desirable from the viewpoint of the self-consistency of the turbulence model. On the other hand, a more simplified evaluation of the local coherent velocity fluctuation, without resorting to the direct numerical calculation of the coherent component of the velocity fluctuation, is preferable from the practical viewpoint for applications of the present model to a stellar convection. In the latter case, the equation of each component of fluctuation energy, (\ref{eq:coh_en_eq}) and (\ref{eq:incoh_en_eq}), with the energy conversion between the coherent and incoherent fluctuations, (\ref{eq:prod_coh_incoh_en}), should provide an important basis of the argument. These explorations are left for interesting future studies.

\section*{Acknowledgements}

The authors are grateful to the anonymous referee for valuable comments improving the presentation of the manuscript. This work was supported by the Japan Society of the Promotion of Science (JSPS) Grants-in-Aid for Scientific Research JP17H06364, JP18H01212, JP18K03700, JP21H01088, and JP21K03612.
Numerical computations were carried out on Cray XC50 at Center for Computational Astrophysics (CfCA), National Astronomical Observatory of Japan (NAOJ). This research was also supported by MEXT as ``Program for Promoting
researches on the Supercomputer Fugaku'' (Toward a unified view of
he universe: from large scale structures to planets, JPMXP1020200109)
and JICFuS. The National Institutes of Natural Sciences
(NINS) program for cross-disciplinary
study (Grant Numbers 01321802 and 01311904) on Turbulence, Transport,
and Heating Dynamics in Laboratory and Solar/Astrophysical Plasmas:
``SoLaBo-X'' supports this work.

\section*{Data Availability}

The data from the numerical simulations and analysis presented in this paper are available from the corresponding author upon reasonable request.



\bibliographystyle{mnras}




\appendix

\section{Turbulent correlations and equations of turbulent statistical quantities}\label{sec:append_A}

From the equations of the fluctuation fields (\ref{eq:fluct_rho_eq})-(\ref{eq:fluct_int_en_eq}), we obtain the equations of turbulent statistical quantities. They are constituted by those of 

\noindent
the turbulent energy:
\begin{Eqnarray}
    \frac{DK}{Dt}
    &=&  - \langle {u'{}^i u'{}^j} \rangle 
    	\frac{\partial U^i}{\partial x^j}
	- (\gamma - 1) \frac{1}{\overline{\rho}} 
    	\langle {e' {\bf{u}}'} \rangle 
    	{\boldsymbol{\cdot}} {\boldsymbol{\nabla}} \overline{\rho}
	- \frac{1}{\overline{\rho}} 
    	\langle {\rho' {\bf{u}}'} \rangle
    	{\boldsymbol{\cdot}} \frac{D{\bf{U}}}{Dt}
	\nonumber\\
	&&+ \frac{1}{\overline{\rho}}
        \langle {\rho' {\textbf{u}}'} \rangle 
        {\boldsymbol{\cdot}} {\textbf{g}}
    - \varepsilon
    + {\boldsymbol{\nabla}} {\boldsymbol{\cdot}} \left( {
        \frac{\nu_{\textrm{T}}}{\sigma_K} {\boldsymbol{\nabla}} K
      } \right),
    \label{eq:model_K_eq}
\end{Eqnarray}
its dissipation rate:
\begin{Eqnarray}
    \frac{D \varepsilon}{Dt}
    &=& - C_{\varepsilon 1} \frac{\varepsilon}{K}
      \langle {u'{}^i u'{}^j} \rangle 
        \frac{\partial U^i}{\partial x^j}
    - C_{\varepsilon 0} \frac{\varepsilon}{K}
      (\gamma - 1) \frac{1}{\overline{\rho}} 
      \langle {e' {\bf{u}}'} \rangle 
      {\boldsymbol{\cdot}} {\boldsymbol{\nabla}} \overline{\rho}
    \nonumber\\
    &&- C_{\varepsilon D} \frac{\varepsilon}{K}
      \frac{1}{\overline{\rho}} 
      \langle {\rho' {\bf{u}}'} \rangle
      {\boldsymbol{\cdot}} \frac{D{\bf{U}}}{Dt}
    + C_{\varepsilon_g} 
        \langle {\rho' {\textbf{u}}'} \rangle 
        	{\boldsymbol{\cdot}} {\textbf{g}}
	\nonumber\\
	&&- C_{\varepsilon 2} \frac{\varepsilon}{K}
      \varepsilon_K
    + {\boldsymbol{\nabla}} {\boldsymbol{\cdot}} \left( {
        \frac{\nu_{\textrm{T}}}{\sigma_\varepsilon} 
        {\boldsymbol{\nabla}} \varepsilon
      } \right),
    \label{eq:model_eps_eq}
\end{Eqnarray}
and the density variance:
\begin{equation}
    \frac{DK_\rho}{Dt}
    = - 2 \left\langle {\rho' {\textbf{u}}'} \right\rangle
      {\boldsymbol{\cdot}} {\boldsymbol{\nabla}} \overline{\rho}
    - 2 K_\rho {\boldsymbol{\nabla}} {\boldsymbol{\cdot}} {\textbf{U}}
    - 2 \left\langle {
        \rho' {\boldsymbol{\nabla}} {\boldsymbol{\cdot}} {\textbf{u}}'
      } \right\rangle \overline{\rho}.
    \label{eq:model_den_var_eq}
\end{equation}

In the mean-field equations (\ref{eq:mean_den_eq})-(\ref{eq:mean_int_en_eq}) and the turbulent statistical quantity equations (\ref{eq:model_K_eq})-(\ref{eq:model_den_var_eq}), we have several turbulent correlations, which include the turbulent mass flux, the Reynolds stress, the turbulent internal-energy flux, etc. Their expressions are explored by analysis of the fluctuation-field equations (\ref{eq:fluct_rho_eq})-(\ref{eq:fluct_int_en_eq}). On the basis of theoretical analysis, the model expressions of the turbulent correlations are given as
\begin{equation}
    \langle {u'{}^i u'{}^j} \rangle_{\textrm{D}}
    \left( {\equiv \langle {u'{}^i u'{}^j} \rangle
    - \frac{1}{3} \langle {{\textbf{u}}'{}^2} \rangle
    \delta^{ij} } \right)
    = - \nu_{\textrm{T}} {\cal{S}}^{ij},
    \label{eq:model_uu_exp}
\end{equation}
\begin{equation}
    \langle {\rho' {\textbf{u}}'} \rangle
    = - \kappa_\rho {\boldsymbol{\nabla}} \overline{\rho}
    - \kappa_E {\boldsymbol{\nabla}} E
    - \kappa_D \frac{D{\textbf{U}}}{Dt},
    \label{eq:model_rho_u_exp}
\end{equation}
\begin{equation}
    \langle {e' {\textbf{u}}'} \rangle
    = - \eta_E {\boldsymbol{\nabla}} E
    - \eta_\rho {\boldsymbol{\nabla}} \overline{\rho},
    \label{eq:model_e_u_exp}
\end{equation}
\begin{equation}
    \langle {e' {\boldsymbol{\nabla}} {\boldsymbol{\cdot}}{\textbf{u}}'}\rangle
    = - \eta_{E1} E - \eta_{E2} \frac{E^2}{\overline{\rho}^2},
    \label{eq:model_e_dil_exp}
\end{equation}
\begin{equation}
    \langle {\rho' {\boldsymbol{\nabla}} {\boldsymbol{\cdot}}{\textbf{u}}'} \rangle
    = \kappa_{E1} \frac{E}{\overline{\rho}}
    - \kappa_{\rho 1} \overline{\rho},
    \label{eq:model_rho_dil_exp}
\end{equation}
where ${\cal{A}}^{ij}_{\textrm{D}} (= {\cal{A}}^{ij} - {\cal{A}}^{\ell\ell} \delta^{ij}/3)$ is the deviatoric or traceless part of tensor $\mbox{\boldmath${\cal{A}}$}$ \citep{yok2018a,yok2018b}. In the presence of rotation and magnetic field, reflectional- or mirror-symmetry is broken. In such cases, several kinds of helicities (kinetic helicity (velocity--vorticity correlation), current helicity (magnetic-field--electric current correlation), cross helicity (velocity--magnetic-field correlation), etc.\ show up in the expressions of these turbulent correlations \citep{yok2013}.

	In (\ref{eq:model_uu_exp})-(\ref{eq:model_rho_dil_exp}), $\nu_{\textrm{T}}$, $\kappa_\rho$, $\kappa_E$, $\kappa_D$, $\eta_E$, etc.\ are the transport coefficients, whose analytical and model expressions are given below in (\ref{eq:kappa_pho_exp_model})-(\ref{eq:kappa_rho1_exp_model}). Among these transport coefficients, the turbulent or eddy viscosity $\nu_{\textrm{T}}$ in (\ref{eq:model_uu_exp}), $\kappa_\rho$ in (\ref{eq:model_rho_u_exp}) and $\eta_E$ in (\ref{eq:model_e_u_exp}) are of primary importance, since they are the coefficients for the gradient-transport models.

	On the basis of the theoretical expressions for the turbulent correlations with the analytical expressions of the transport coefficients in terms of the temporal and spectral integrals of the Green's functions and spectral functions, we construct a system of turbulence model equations. The transport coefficients in (\ref{eq:model_uu_exp})-(\ref{eq:model_rho_dil_exp}) are modelled with time scales associated with the Green's functions, and one-point turbulence statistical quantities related with the spectral functions. With introducing the abbreviated form of the time and spectral integrals as
\begin{equation}
    I_0 \left\{ {A,B} \right\}
    = \int\!\! d{\bf{k}} \int_{-\infty}^\tau \hspace{-10pt}d\tau_1
        A(k;\tau,\tau_1) B(k;\tau,\tau_1),
    \label{eq:abbrev_int_1}
\end{equation}
\begin{eqnarray}
    &&I_{2n} \left\{ {A^{(1)},B^{(2)},C^{(3)},D^{(3)}} \right\}
    = \int\!\! d{\bf{k}}\ k^{2n}
        \int_{-\infty}^\tau \hspace{-10pt}d\tau_1
        \int_{-\infty}^\tau \hspace{-10pt}d\tau_2
        \int_{-\infty}^\tau \hspace{-10pt}d\tau_3
	\nonumber\\
   &&\hspace{20pt} \rule{0.ex}{3.ex} \times  A(k;\tau,\tau_1) B(k;\tau,\tau_2) 
            C(k;\tau,\tau_3) D(k;\tau_2,\tau_3),
        \label{eq:abbrev_int_2}
\end{eqnarray}
the transport coefficients are expressed and modelled as
\begin{equation}
    \kappa_{\rho}
= \frac{1}{3} I_0 \left\{ {
        G_\rho, 2Q_{u{\rm{S}}} + Q_{u{\rm{C}}}
    } \right\}
    = C_{\kappa\rho} \tau_\rho \langle {{\textbf{u}}'{}^2} \rangle/2,
    \label{eq:kappa_pho_exp_model}
\end{equation}
\begin{equation}
    \kappa_{E}
    = \frac{1}{3} (\gamma - 1) 
        \frac{1}{\overline{\rho}} I_0 \left\{ {
            2 G_{u{\rm{S}}} + G_{u{\rm{C}}}, Q_\rho
        } \right\}
    = C_{\kappa E} (\gamma - 1) \tau_u  
        \overline{\rho} 
        \frac{\langle {\rho'{}^2} \rangle}{\overline{\rho}^2},
    \label{eq:kappa_Q_exp_model}
\end{equation}
\begin{equation}
    \kappa_{D}
    = \frac{1}{3\overline{\rho}} I_0 \left\{ {
        2G_{u{\rm{S}}} + G_{u{\rm{C}}}, Q_\rho
    } \right\}
    = C_{\kappa D} \tau_u
        \overline{\rho} 
        \frac{\langle {\rho'{}^2} \rangle}{\overline{\rho}^2},
    \label{eq:kappa_D_exp_model}
\end{equation}
\begin{Eqnarray}
    \nu_{\textrm{T}}
    &=& \frac{1}{15} \left( {
        7 I_0 \{ {G_{u{\textrm{S}}}, Q_{u{\textrm{S}}}} \}
        + 3 I_0 \{ {G_{u{\textrm{S}}}, Q_{u{\textrm{C}}}} \}
	} \right.
	\nonumber\\
	&&\left. { \hspace{20pt}
        + 3 I_0 \{ {G_{u{\textrm{C}}}, Q_{u{\textrm{S}}}} \}
        + 2 I_0 \{ {G_{u{\textrm{C}}}, Q_{u{\textrm{C}}}} \}
    } \right)
    \nonumber\\
    &=& C_\nu \tau_u \langle {{\textbf{u}}'{}^2} \rangle/2,
    \label{eq:nu_K_exp_model}
\end{Eqnarray}
\begin{equation}
    \eta_E
    = \frac{1}{3} 
    I_0 \{ {G_q, 2Q_{u{\textrm{S}}} + Q_{u{\textrm{C}}}} \}
    = C_{\eta E} \tau_e \langle {{\textbf{u}}'{}^2} \rangle/2,
    \label{eq:eta_Q_exp_model}
\end{equation}
\begin{Eqnarray}
    \eta_{\rho}
    &=& \frac{1}{3} (\gamma - 1) \frac{1}{\overline{\rho}}
        I_0 \left\{ {
        2 G_{u{\rm{S}}} + G_{u{\rm{C}}}, Q_e
    } \right\}
    \nonumber\\
    &=& C_{\eta\rho} (\gamma - 1) \tau_u
        \frac{\langle {e'{}^2} \rangle}{\overline{\rho}}
	\nonumber\\
	&=& C_{\eta\rho} (\gamma - 1)^3 
    \frac{\tau_u \tau_e^2}{\tau_\rho^2} 
    \frac{E^2}{\overline{\rho}}
    \frac{\langle {\rho'{}^2} \rangle}{\overline{\rho}^2},
    \label{eq:eta_rho_exp_model}
\end{Eqnarray}
\begin{equation}
    \eta_{E1}
    = I_1 \{ {G_{u{\textrm{S}}} - G_\rho, Q_{u{\textrm{C}}}} \},
    \label{eq:eta_Q1_exp_model}
\end{equation}
\begin{equation}
    \eta_{E2}
    = (\gamma - 1) 
        I_1 \{ {G_{u{\textrm{C}}}, Q_{\rho}} \},
    \label{eq:eta_Q2_exp_model}
\end{equation}
\begin{equation}
    \kappa_{E1}
    = (\gamma - 1) 
        I_1 \{ {G_{u{\textrm{C}}}, Q_\rho} \},
    \label{eq:kappa_Q1_exp_model}
\end{equation}
\begin{equation}
    \kappa_{\rho 1}
    = I_1 \{ {G_\rho, Q_{u{\textrm{C}}}} \},
    \label{eq:kappa_rho1_exp_model}
\end{equation}
where in the indices ${\textrm{S}}$ and ${\textrm{C}}$ denote the solenoidal and compressible components, respectively. In the final equality of (\ref{eq:eta_rho_exp_model}), use has been made of the approximate relations:
\begin{equation}
    e' \simeq - (\gamma-1) \tau_e E {\boldsymbol{\nabla}} {\boldsymbol{\cdot}} {\textbf{u}}'
    \label{eq:q_turb_dil_rel}
\end{equation}
and
\begin{equation}
    {\boldsymbol{\nabla}} {\boldsymbol{\cdot}} {\textbf{u}}'
    \simeq - \frac{1}{\tau_\rho} \frac{\rho'}{\overline{\rho}}.
    \label{eq:turb_dil_rho_rel}
\end{equation}
Here we used time scales associated with the evolution of density, velocity, and internal energy as
\begin{equation}
    \tau_s = \int_{-\infty}^\tau\!\!\! d\tau_1 
    \langle {G'_s({\textbf{k}};\tau,\tau_1)} \rangle
    = \int_{-\infty}^\tau\!\!\! d\tau_1 G_s({\textbf{k}};\tau,\tau_1)
    \label{eq:timescales_rho_u_q}
\end{equation}
with $s=(\rho,u,e)$.

\section{Derivation of evolution equations of the coherent and incoherent fluctuation energies}\label{sec:append_B}

	We derive the evolution equation of the Reynolds stress from the equation of the velocity fluctuation. In convective turbulence, the compressibility effect represented by the density fluctuation $\rho'$, which is connected to the mean density stratification $\nabla \langle {\rho} \rangle$ and the turbulent dilatation $\nabla \cdot {\bf{u}}'$, plays an important role. In this sense, we have to treat the compressible velocity fluctuation equation. However, in order to focus on the effect of the double-averaging procedure, we confine ourselves to the simplest solenoidal velocity fluctuation equation in this Appendix~\ref{sec:append_B} as
\begin{Eqnarray}
	\frac{\partial u'{}^i}{\partial t}
	&+& \langle {u} \rangle ^\ell 
	\frac{\partial u'{}^i}{\partial x^\ell}
	= - u'{}^\ell 
	\frac{\partial \langle {u} \rangle^i}{\partial x^\ell}
	- \frac{1}{\rho_0} \frac{\partial p'}{\partial x^i}
	\nonumber\\
	&-& \frac{\partial}{\partial x^\ell} \left( {
		u'{}^\ell u'{}^i 
		- \langle {u'{}^\ell u'{}^i} \rangle
    } \right)
	+ \nu \frac{\partial^2 u'{}^i}{\partial x^\ell \partial x^\ell}.
	\label{eq:fluct_u_eq}
\end{Eqnarray}
An extension to the compressible case requires a cumbersome but straightforward calculation. As for the velocity fluctuation equations in the strongly compressible magnetohydrodynamic (MHD) case, the reader is referred to \citet{yok2018a,yok2018b}.

\subsection{Evolution equations of the coherent and incoherent Reynolds stresses}
We multiply $u'{}^j$ on (\ref{eq:fluct_u_eq}) and take the time--space double averaging. With considering the relations (\ref{eq:time-space_ave_rels}), we decompose a field quantity as (\ref{eq:double_ave_decomp}). Exchanging $i$ and $j$ and adding them, we obtain the equations of the dispersion or coherent component and random or incoherent component of the Reynolds stress. They are given as
\begin{eqnarray}
	\lefteqn{
	\left( {
		\frac{\partial}{\partial t}
		+ \langle {u} \rangle ^\ell 
		\frac{\partial} {\partial x^\ell}
	} \right) \langle {\tilde{u}^i \tilde{u}^j} \rangle
	}\nonumber\\
	&&\hspace{0pt}
	= - \langle {\tilde{u}^j \tilde{u}^\ell} \rangle 
		\frac{\partial \langle {u} \rangle^i}{\partial x^\ell}
	- \langle {\tilde{u}^i \tilde{u}^\ell} \rangle 
		\frac{\partial \langle {u} \rangle^j}{\partial x^\ell}
	\nonumber\\
	&& - \frac{1}{\rho_0} \left\langle {
		\tilde{p} \left( {
			\frac{\partial \tilde{u}^i}{\partial x^j}
			+ \frac{\partial \tilde{u}^j}{\partial x^i}
		} \right)
	} \right\rangle
	- 2 \nu \left\langle {
		\frac{\partial \tilde{u}^i}{\partial x^\ell} 
		\frac{\partial \tilde{u}^j}{\partial x^\ell}
	} \right\rangle
	\nonumber\\
	&&+ \frac{\partial}{\partial x^\ell} 
		\left( {
			- \left\langle {
				\tilde{u}^\ell \tilde{u}^i \tilde{u}^j
			} \right\rangle
		+ \langle {\tilde{p} \tilde{u}^i} \rangle \delta^{\ell j}
    	+ \langle {\tilde{p} \tilde{u}^j} \rangle \delta^{\ell i}
    	\rule{0.ex}{3.0ex}
  		+ \nu \frac{\partial}{\partial x^\ell}
			\langle {\tilde{u}^i \tilde{u}^j} \rangle
		} \right)
	\nonumber\\
	&&+ \left\langle {
		\widetilde{u''{}^\ell u''{}^j}
		\frac{\partial \tilde{u}^i}{\partial x^\ell}
	} \right\rangle
	+ \left\langle {
		\widetilde{u''{}^\ell u''{}^i}
		\frac{\partial \tilde{u}^j}{\partial x^\ell}
	} \right\rangle
	\nonumber\\
	&&- \frac{\partial}{\partial x^\ell} 
		\left\langle {
			\widetilde{u''{}^\ell u''{}^i} \tilde{u}^j
			+ \widetilde{u''{}^\ell u''{}^j} \tilde{u}^i
		} \right\rangle,
	\label{eq:coh_rey_strss_eq}
\end{eqnarray}
\begin{eqnarray}
	\lefteqn{
	\left( {
		\frac{\partial}{\partial t}
		+ \langle {u} \rangle ^\ell 
		\frac{\partial} {\partial x^\ell}
	} \right) \langle {u''{}^i u''{}^j} \rangle
	}\nonumber\\
	&&\hspace{0pt}
	= - \langle {u''{}^j u''{}^\ell} \rangle 
		\frac{\partial \langle {u} \rangle^i}{\partial x^\ell}
	- \langle {u''{}^i u''{}^\ell} \rangle 
		\frac{\partial \langle {u} \rangle^j}{\partial x^\ell}
	\nonumber\\
	&& - \frac{1}{\rho_0} \left\langle {
		p'' \left( {
			\frac{\partial u''{}^i}{\partial x^j}
			+ \frac{\partial u''{}^j}{\partial x^i}
		} \right)
	} \right\rangle
	- 2 \nu \left\langle {
		\frac{\partial u''{}^i}{\partial x^\ell} 
		\frac{\partial u''{}^j}{\partial x^\ell}
	} \right\rangle
	\nonumber\\
	&&+ \frac{\partial}{\partial x^\ell} 
		\left( {
	- \left\langle {
		u''{}^\ell u''{}^i u''{}^j
	} \right\rangle
    + \langle {p'' u''{}^i} \rangle \delta^{\ell j}
	+ \langle {p'' u''{}^j} \rangle \delta^{\ell i}
    \rule{0.ex}{3.0ex}
	} \right.
	\nonumber\\
	&&\hspace{35pt}\left. {
	+ \nu \frac{\partial}{\partial x^\ell}
		\langle {u''{}^i u''{}^j} \rangle
	} \right)
	\nonumber\\
	&&- \left\langle {
		\widetilde{u''{}^\ell u''{}^j}
		\frac{\partial \tilde{u}^i}{\partial x^\ell}
	} \right\rangle
	- \left\langle {
		\widetilde{u''{}^\ell u''{}^i}
		\frac{\partial \tilde{u}^j}{\partial x^\ell}
	} \right\rangle
	\nonumber\\
	&&- \frac{\partial}{\partial x^\ell} 
	\left\langle {
		\widetilde{u''{}^i u''{}^j} \tilde{u}^\ell
	} \right\rangle.
	\label{eq:incoh_rey_strss_eq}
\end{eqnarray}
	In each equation, the first and second terms represent the production of the stress due to the mean velocity gradients. The third term is the re-distribution due to the pressure--strain, and the fourth term represents
the viscous dissipation. The fifth term represents the transport expressed in the divergence form. The final three terms are terms newly appeared due to the double averaging procedure. Among these newly appeared terms, the first and second terms are production due to the gradients of the coherent fluctuations, and the third term, expressed in the divergence form, is the transport term. The $\widetilde{{\bf{u}}'' {\bf{u}}''}$ in these three terms are defined by
\begin{eqnarray}
	\lefteqn{
	\widetilde{u''{}^i u''{}^j}
	= \overline{u''{}^i u''{}^j}
	- \langle {\overline{{u''{}^i u''{}^j}}} \rangle
	}\nonumber\\
	&&\hspace{13pt}= \overline{u''{}^i u''{}^j}
	- \langle {{u''{}^i u''{}^j}} \rangle,
	\label{eq:disp_incoh_strss}
\end{eqnarray}
which is the dispersion part of the random/incoherent correlation, defined by the time correlation of the random/incoherent fluctuations with the space correlation part being subtracted.

\subsection{Evolution equations of the coherent and incoherent energies}
Taking the contraction of $i$ and $j$ in (\ref{eq:coh_rey_strss_eq}) and (\ref{eq:incoh_rey_strss_eq}), we obtain the evolution equations of the coherent- and incoherent-fluctuation energies as
\begin{eqnarray}
	\lefteqn{
	\left( {
		\frac{\partial}{\partial t}
		+ \langle {u} \rangle ^\ell \frac{\partial} {\partial x^\ell}
	} \right) \left\langle {\frac{1}{2}(\tilde{u}^j)^2} \right\rangle
	}\nonumber\\
	&&= - \langle {\tilde{u}^j \tilde{u}^\ell} \rangle 
			\frac{\partial \langle {u} \rangle^j}{\partial x^\ell}
		- \frac{1}{\rho_0} \left\langle {
			\tilde{p} \left( {
			\frac{\partial \tilde{u}^j}{\partial x^j}
		} \right)
	} \right\rangle
	- \nu \left\langle { \left( {
		\frac{\partial \tilde{u}^j}{\partial x^\ell} 
	} \right)^2
	} \right\rangle
	\nonumber\\
	&&+ \frac{\partial}{\partial x^\ell} 
	\left\langle {
		- \tilde{u}^\ell \frac{1}{2} (\tilde{u}^j)^2
		+ \tilde{p} \tilde{u}^\ell
		+ \nu \frac{\partial}{\partial x^\ell}
			\frac{1}{2} (\tilde{u}^j)^2
	} \right\rangle
	\nonumber\\
	&&+ \left\langle {
		\widetilde{u''{}^\ell u''{}^j}
		\frac{\partial \tilde{u}^j}{\partial x^\ell}
	} \right\rangle
	- \frac{\partial}{\partial x^\ell} 
		\left\langle {
		\widetilde{u''{}^\ell u''{}^j} \tilde{u}^j
	} \right\rangle,
	\label{eq:coh_en_eq}
\end{eqnarray}
\begin{eqnarray}
	\lefteqn{
	\left( {
		\frac{\partial}{\partial t}
		+ \langle {u} \rangle ^\ell \frac{\partial} {\partial x^\ell}
	} \right) 
	\left\langle {\frac{1}{2} (u''{}^j)^2} \right\rangle
	}\nonumber\\
	&&= - \langle {u''{}^j u''{}^\ell} \rangle 
	\frac{\partial \langle {u} \rangle^j}{\partial x^\ell}
	- \frac{1}{\rho_0} \left\langle {
		p'' \left( {\frac{\partial u''{}^j}{\partial x^j}} \right)
	} \right\rangle
	- \nu \left\langle {
		\left( {\frac{\partial u''{}^j}{\partial x^\ell}} \right)^2 
	} \right\rangle
	\nonumber\\
	&&+ \frac{\partial}{\partial x^\ell} 
		\left\langle {
			- u''{}^\ell \frac{1}{2} (u''{}^j)^2
    		+ p'' u''{}^\ell
			+ \nu \frac{\partial}{\partial x^\ell} \frac{1}{2} (u''{}^j)^2
		} \right\rangle
	\nonumber\\
	&&- \left\langle {
		\widetilde{u''{}^\ell u''{}^j}
		\frac{\partial \tilde{u}^j}{\partial x^\ell}
	} \right\rangle
	- \frac{\partial}{\partial x^\ell} 
		\left\langle {
			\widetilde{\frac{1}{2} (u''{}^j)^2} \tilde{u}^\ell
		} \right\rangle.
	\label{eq:incoh_en_eq}
\end{eqnarray}
Equations~(\ref{eq:coh_en_eq}) and (\ref{eq:incoh_en_eq}) provide a clear picture on the property of energy transfer in convective turbulence. The first term on the right-hand-side of each of (\ref{eq:coh_en_eq}) and (\ref{eq:incoh_en_eq}) represents the production of the energy arising from the mean velocity shear, the second term is the pressure--dilatation, which vanishes in the solenoidal case. The third term is the dissipation rates due to the viscosity, and the fourth term written in the divergence form is the transport rate representing the flux through the boundary. All these terms are in the same form as the equation of the total fluctuation energy $\langle {{\bf{u}}'{}^2} \rangle/2$.

\section{Simple plume equations}\label{sec:append_C}
In this work, we regard plume motion as a coherent component of fluctuations, whose statistical property is a key to evaluate the non-equilibrium effect through $\tilde{\varepsilon}$ in (\ref{eq:turb_mass_flux_kappa_noneq}). In order to evaluate the statistical properties of coherent fluctuations, represented by the fluctuation energy, its dissipation rate, time and length scales, etc., we utilise an argument on the plume dynamics. In this Appendix~\ref{sec:append_C}, following \citet{tur1973} and \citet{lin2000}, we present only some basics of plume dynamics which are relevant to the evaluation of the coherent fluctuation energy and its time and length scales. Hereafter, a plume quantity $\tilde{f}$ will be denoted as $f$ with the tilde symbol being suppressed.

	We consider axisymmetric steady flow of an inviscid incompressible fluid with no swirl. The velocity is written as ${\bf{u}} = (u^r, 0, u^z)$ in cylindrical coordinate $(r, \theta, z)$ with $z$ being vertical. The flow is governed by the continuity equation:
\begin{equation}
	\frac{1}{r} \frac{\partial}{\partial r} r u^r
	+ \frac{\partial u^z}{\partial z}
	= 0,
	\label{eq:cont_eq}
\end{equation} 
and the equations of the radial and vertical velocity components, $u^r$ and $u^z$, as
\begin{equation}
	u^r \frac{\partial u^r}{\partial r}
	+ u^z \frac{\partial u^r}{\partial z}
	= - \frac{1}{\rho_0} \frac{\partial p}{\partial r},
	\label{eq:ur_eq}
\end{equation}
\begin{equation}
	u^r \frac{\partial u^z}{\partial r}
	+ u^z \frac{\partial u^z}{\partial z}
	= - \frac{1}{\rho_0} \frac{\partial p}{\partial z}
	- g \frac{\rho}{\rho_0},
	\label{eq:uz_eq}
\end{equation}
where $\rho_0$ is a reference density, which can be represented by the average density for the Boussinesq convection. In (\ref{eq:ur_eq})-(\ref{eq:uz_eq}), the turbulence correlation terms such as $\langle {u^r u^z} \rangle$, $\langle {u^r u^z} \rangle$, etc.\ were dropped since their coupling with the fluctuating fields $u^r$, $u^z$, etc.\ are expected to be negligibly small in the energy equation. Namely, this approximation is not bad for treating energetics of fluctuating motions. In this sense, we can follow the ``laminar'' argument of the plume dynamics based on (\ref{eq:cont_eq})-(\ref{eq:uz_eq}).

	There are some quantities that characterise dynamics of plume. Among them, the volume flux $Q$, the momentum flux $M$, and the buoyancy flux $B$ are of primary importance. They are defined as
\begin{equation}
	Q \equiv 2\pi \int_0^\infty r u^z dr,
	\label{eq:volume_flux_def}
\end{equation}
\begin{equation}
	M \equiv 2\pi \int_0^\infty r u^z u^z dr,
	\label{eq:mt_flux_def}
\end{equation}
\begin{equation}
	B \equiv 2\pi \int_0^\infty r u^z g 
		\left( {\frac{\rho_e - \rho}{\rho_0}} \right) dr,
	\label{eq:buoy_flux_def}
\end{equation}
where $\rho$ is the density of plume and $\rho_e$ is the density of environment. For a Boussinesq plume, the buoyancy flux $B$ is conserved. In  more general unstratified fluid cases for the non-Boussinesq plume, the flux of weight deficiency defined by
\begin{equation}
	\frac{\rho B}{g} 
	= 2\pi \int_0^\infty r u^z g \left( {\rho_e - \rho} \right) dr 
	\label{eq:weight_deficiency_def}
\end{equation}
is known to be conserved \citep{roo1996}.

	If we adopt the simplest possible ``top-hat'' profiles for velocity and buoyancy, the volume and momentum fluxes are given as
\begin{equation}
	Q = 2\pi \int_0^\infty u^z r dr = \pi \hat{b}^2 \hat{w},
	\label{eq:tophat_Q_exp}
\end{equation}
\begin{equation}
	M = 2\pi \int_0^\infty (u^z)^2 r dr = \pi \hat{b}^2 \hat{w}^2
	\label{eq:tophat_M_exp}
\end{equation}
with $\hat{w}$ and $\hat{b}$ being the top-hat velocity and radius of the plume. Conversely, the top-hat variables $\hat{w}$ and $\hat{b}$ are expressed in terms of $Q$ and $M$ as
\begin{equation}
	\hat{w} = \frac{M}{Q},
	\label{eq:tophat_w_exp}
\end{equation}
\begin{equation}
	\hat{b} = \frac{Q}{\pi^{1/2}M^{1/2}}.
	\label{eq:tophat_b_exp}
\end{equation}

	In order to abstract basic properties of plume dynamics, it is useful to consider a cone-shaped self-similar plume originating from a point source in a stationary ambient fluid. In such a self-similar plume, the properties of plume depend on the buoyancy flux $B$ and the height or depth $z$ only. Dimensional analysis shows at each height/depth, the mean vertical velocity $u^z$, the mean buoyancy $g' [\equiv g (\rho_e - \rho)/\rho_0]$ and the mean radius of the plume $b$ are given by
\begin{equation}
	u^z = c_z B^{1/3} z ^{-1/3},
	\label{eq:uz_scaling}
\end{equation}
\begin{equation}
	g' \equiv g \frac{\rho_e - \rho}{\rho_0}
	= c_g B^{2/3} z ^{-5/3},
	\label{ea:g_scaling}
\end{equation}
\begin{equation}
	b = \beta z,
	\label{eq:b_scaling}
\end{equation}
respectively. Here, $\beta$ is a dimensionless constant, and $c_z$ and $c_g$ are dimensionless functions of the scaled radius $r/b$.

	For a self-similar plume (\ref{eq:uz_scaling})-(\ref{eq:b_scaling}), the volume flux $Q$ [defined by (\ref{eq:volume_flux_def})] is given by
\begin{equation}
	Q = \pi c_Q \beta^2 B^{1/3} z^{5/3}
	\label{eq:Q_B_scaling}
\end{equation}
[$c_Q$: dimensionless constant] \citep{lin2000}. This shows that the volume flux $Q$ increases with $B$ and $z$ due to the entrainment of ambient fluid into the plume.

	On the other hand, if we multiply (\ref{eq:uz_eq}) by any power of the vertical velocity, $(u^z)^n$, and integrate across the plume with respect to $r$ from $r = 0$ to $r = \infty$, we have
\begin{equation}
	2\pi \int_0^\infty (u^z)^n r \left( {
		u^r \frac{\partial u^z}{\partial r}
		+ u^z \frac{\partial u^z}{\partial z}
	} \right) dr
	= 2\pi \int_0^\infty (u^z)^n 
  \frac{\rho_{\textrm{e}} - \rho}{\rho_0} g r dr.
  \label{eq:n_moment_eq}
\end{equation}
Noting that
\begin{equation}
	(u^z)^n \frac{\partial u^z}{\partial r}
	= \frac{1}{n+1} \frac{\partial}{\partial r} (u^z)^{n+1},
	\label{eq:r_deriv_rule}
\end{equation}
\begin{equation}
	(u^z)^n \frac{\partial u^z}{\partial z}
	= \frac{1}{n+1} \frac{\partial}{\partial z} (u^z)^{n+1},
	\label{eq:z_deriv_rule}
\end{equation}
the left-hand-side of (\ref{eq:n_moment_eq}) can be integrated by parts as
\begin{Eqnarray}
	\int_0^\infty r u^r \frac{\partial}{\partial r} (u^z)^{n+1} dr
	&=& \left[ {r u^r (u^z)^{n+1}} \right]_0^\infty
	- \int_0^\infty \left( {
		\frac{\partial}{\partial r} r u^r
	} \right) (u^z)^{n+1} dr
	\nonumber\\
	&=& - \int_0^\infty \left( {
		- r \frac{\partial u^z}{\partial z}
	} \right) (u^z)^{n+1} dr.
	\label{eq:int_by_parts}
\end{Eqnarray}
Here, use has been made of $u^z = 0$ outside of the plume, and made of the continuity equation (\ref{eq:cont_eq}).
Using (\ref{eq:int_by_parts}), (\ref{eq:n_moment_eq}) is written as
\begin{equation}
	\frac{2\pi}{n+1} \frac{d}{dz} \int_0^\infty r (u^z)^{n+2} dr
	= 2\pi \int_0^\infty (u^z)^n 
  		\frac{\rho_{\textrm{e}} - \rho}{\rho_0} g r dr.
	\label{eq:mt_rho_rel_n}
\end{equation}
Note that the numerical factor in the present result in (\ref{eq:mt_rho_rel_n}) is different from the counterpart in \citet{lin2000}.

	Putting $n=0$ in (\ref{eq:mt_rho_rel_n}), we have
\begin{equation}
	2\pi \frac{d}{dz} \int_0^\infty r (u^z)^2 dr
	\equiv \frac{dM}{dz}
	= 2\pi g \int_0^\infty 
		\frac{\rho_{\textrm{e}} - \rho}{\rho_0} r dr.
	\label{eq:mt_rho_rel_0}
\end{equation}
This suggests that the momentum flux $M$ in the plume is changed by the buoyancy force.
	Putting $n=1$ in (\ref{eq:mt_rho_rel_n}), we have
\begin{equation}
	\pi \frac{d}{dz} \int_0^\infty r (u^z)^3 dr
	= 2\pi g \int_0^\infty u^z 
		\frac{\rho_{\textrm{e}} - \rho}{\rho_0} r dr.
	\label{eq:mt_rho_rel_1}
\end{equation}
	The vertical flux of kinetic energy is expressed as
\begin{equation}
	2\pi \frac{d}{dz} \int_0^\infty r u^z (u^z)^2 dr
	= 2B.
	\label{eq:en_flux_buoy_rel}
\end{equation}
This shows that the vertical flux of the kinetic energy changes due to the buoyancy flux $B$.


\bsp	
\label{lastpage}
\end{document}